\newif\ifsubmode
\submodefalse

\newif\ifprintfig
\printfigfalse

\ifsubmode
  \documentstyle[12pt,aasms4,epsf]{article}
  \received{}
  \accepted{}
  \journalid{}{}
  \articleid{}{}
\else
  \documentstyle[11pt,aaspp4,epsf]{article}
  \slugcomment{{\sl submitted to The Astrophysical Journal; Feb 17, 1997.}}
\fi

\lefthead{van der Marel, de Zeeuw \& Rix}
\righthead{A black hole in M32 --- I.~HST/FOS Observations}

\def\etal{{et al.~}}
\def\lta{\lesssim}
\def\gta{\gtrsim}
\def\kms{\>{\rm km}\,{\rm s}^{-1}}

\def\Msun{\>{\rm M_{\odot}}}
\def\fr#1#2{{\textstyle {#1 \over #2}}}

\def\tsigma{{\tilde \sigma}}

\def\refsAppA{A}
\def\reftobsetup{2}
\def\reftapplaces{3}
\def\reftpsfapfit{5}
\def\refftransmis{8}

\begin{document}

\title{Improved evidence for a black hole in M32 from HST/FOS spectra ---\\
       I.~Observations\altaffilmark{1}}

\author{Roeland P.~van der Marel\altaffilmark{2}}
\affil{Institute for Advanced Study, Olden Lane, Princeton, NJ 08540}

\author{P.~Tim de Zeeuw}
\affil{Sterrewacht Leiden, Postbus 9513, 2300 RA Leiden, The Netherlands}

\author{Hans--Walter Rix\altaffilmark{3}}
\affil{Steward Observatory, University of Arizona, Tucson, AZ 85721}


\altaffiltext{1}{Based on observations with the NASA/ESA Hubble Space 
       Telescope obtained at the Space Telescope Science Institute, which is 
       operated by the Association of Universities for Research in Astronomy, 
       Incorporated, under NASA contract NAS5-26555.}

\altaffiltext{2}{Hubble Fellow.}

\altaffiltext{3}{Alfred P. Sloan Fellow.}


\ifsubmode\else
\clearpage\fi


\ifsubmode\else
\baselineskip=14pt
\fi


\begin{abstract}
We have obtained spectra through small apertures centered on the
nuclear region and major axis of M32, with the Faint Object
Spectrograph (FOS) on the Hubble Space Telescope (HST). A detailed
analysis and reduction of the data is presented, including: (i) new
calibrations and modeling of the FOS aperture sizes,
point-spread-function and line-spread-functions; (ii) determination of
the aperture positioning for each observation from the observed count
rate; and (iii) accurate wavelength calibration, template matching,
and kinematical analysis of the spectra. This yields measurements of
the stellar rotation velocities and velocity dispersions near the
center of M32, with five times higher spatial resolution than the best
available ground-based data. The inferred velocities provide the
highest angular resolution stellar kinematical data obtained to date
for any stellar system.

The HST observations show a steeper rotation curve and higher central
velocity dispersion than the ground-based data. The rotation velocity
is observed to be $\sim 30 \kms$ at $0.1''$ from the nucleus. This is
roughly twice the value measured from the ground at this distance. The
nuclear dispersion measured through the smallest FOS aperture
($0.068''$ square) is $156 \pm 10 \kms$. The average of four
independent dispersion measurements at various positions inside the
central $0.1''$ is $126 \kms$, with a RMS scatter of $21 \kms$. The
nuclear dispersion measured from the ground is only 85--95$\kms$,
whereas the dispersion outside the central arcsec is only $\sim
45$--$55 \kms$. These results significantly strengthen previous
arguments for the presence of a massive nuclear black hole in
M32. Detailed dynamical models are presented in a series of companion
papers.
\end{abstract}



\clearpage


\section{Introduction}

Astronomers have sought for two decades for dynamical evidence for the
presence of massive black holes in galaxies by studying the dynamics
of gas and stars in their nuclei. Rapid motions provide the main
signature of a black hole. If these are indeed observed, the main
difficulty is to rule out alternative interpretations. Gas motions can
be due to hydrodynamical processes (inflow, outflow, turbulence, etc.)
in addition to gravity. Large stellar velocities in galactic nuclei
can be the result of an overabundance of stars on radial orbits. The
primary observational tool to discriminate between different models is
to obtain data of the highest possible spatial resolution.

In recent years much progress has been made in all areas of this
field. For active galaxies, the existence of a central dark mass in at
least some galaxies is now well established. The rotation velocities
of nuclear gas disks detected with the HST can be used to study the
central mass distribution. This has yielded evidence for a central
dark mass of $2.4 \times 10^9 \Msun$ in M87 (Ford \etal 1994; Harms
\etal 1994), and of $5 \times 10^8 \Msun$ in NGC 4261 (Ferrarese, 
Ford \& Jaffe 1996). Even higher spatial resolution VLBA radio
observations of water masers in the nucleus of the active galaxy NGC
4258 (Miyoshi \etal 1995) have revealed a torus in Keplerian rotation
around a dark mass of $3.6 \times 10^7 \Msun$. For the case of NGC
4258, there are strong theoretical arguments that this mass is indeed
a black hole (Maoz 1995).

The density of quasars at high redshifts suggests that many currently
normal galaxies had an active phase in the past (Chokshi \& Turner
1992; Haehnelt \& Rees 1993). Hence, black holes are believed to be
common in quiescent galaxies as well. In these galaxies only stellar
kinematics are generally available to study the nuclear mass
distribution. The evidence for a black hole in our own Galaxy is now
very strong, due to proper motion measurements for individual stars
near Sgr~A$^*$ (Eckart \& Genzel 1996, 1997). For a handful of other,
nearby galaxies, evidence for a central dark mass was obtained from
ground-based measurements of line-of-sight velocities (see Kormendy \&
Richstone 1995 for a review), but it remained difficult to rule out
all alternative models. It was always foreseen to be a main task for
the HST to improve the evidence, by providing spectra of superior
spatial resolution. Stellar kinematical studies with HST became
possible after the refurbishment mission in 1993. The first results
were presented by Kormendy \etal (1996a,b), for NGC 3115 and NGC
4594. Spectra near the nucleus with $\sim 0.2''$ resolution confirmed
previous arguments for black holes of $2 \times 10^9 \Msun$ and $1
\times 10^9 \Msun$, respectively.

The quiescent E3 galaxy M32 has long been one of the best-studied
black hole candidate galaxies. The presence of a central dark mass of
$10^6$--$10^7 \Msun$ has been argued on the basis of ground-based data
with continuously increasing spatial resolution. The most recent work
indicates a central dark mass of $2$--$3 \times 10^6 \Msun$ (van der
Marel et al.~1994b; Qian et al.~1995; Dehnen 1995; Bender, Kormendy \&
Dehnen 1996). However, the best ground-based kinematical data still
have a spatial resolution of `only' $\sim 0.5''$. Goodman \& Lee
(1989) showed that this is insufficient to rule out a cluster of dark
objects, as opposed to a central black hole, on the basis of
theoretical arguments. In addition, none of the previous dynamical
studies has considered axisymmetric stellar dynamical models with
fully general phase-space distribution functions. Hence, it has not
been shown convincingly that no plausible model can be constructed
that fits the data without requiring a central dark mass. Higher
spatial resolution data are therefore highly desirable.

This paper is part of a series in which we present the first HST
spectra and new dynamical models for M32. We obtained spectra of the
nuclear region with the HST/FOS, and we give here a detailed
description of the acquisition and analysis of these data. Dynamical
models are presented in van der Marel \etal (1997b) and Cretton \etal
(1997). The main results of the project are summarized and discussed
in van der Marel \etal (1997a).

The instrumental resolution of the FOS has a Gaussian dispersion of
$\sim 100 \kms$, while the wavelength scale can vary at the $20 \kms$
level from orbit to orbit. A study of a low mass galaxy such as M32
(the main body of which has a velocity dispersion of $\sim 50 \kms$
and a rotation velocity amplitude of $\sim 45 \kms$) is therefore
significantly more complicated observationally, than that of more
massive galaxies (which have higher dispersions). An additional
complication is that the `sphere of influence' of the suspected black
hole in M32 is smaller than that of most other candidate galaxies. It
is thus necessary to use the smallest FOS apertures, for which it is
more difficult to do an accurate target acquisition. In addition, for
a proper interpretation of the results it is necessary to determine
the aperture position for each observation {\it post facto\/}, with an
accuracy of $\sim 0.01''$. To deal with these complications it proved
necessary to study the instrument and analyze the data in more than
the usual detail. Parts of this paper are therefore of a somewhat
technical nature. Readers interested mostly in the stellar kinematical
results may wish to skip directly to Section~\ref{ss:results}.

The paper is organized as follows. Section~\ref{s:obsetup} summarizes
the observational setup and strategy. Section~\ref{s:pos} discusses
the aperture positions for the observations. Section~\ref{s:datareduc}
describes some aspects of the data reduction. Section~\ref{s:lsf}
presents calculations of the line-spread function for each
observation. Section~\ref{s:temp} discusses the template spectrum used
in modeling the M32 spectra. Section~\ref{s:kinematics} discusses the
kinematical analysis. Section~\ref{s:discussion} summarizes the main
observational results. In the appendices, new calibrations and
modeling are presented of the FOS aperture sizes,
point-spread-function (PSF) and line-spread-functions, for which no
sufficiently accurate determinations were previously available.

\section{Observational setup and strategy}
\label{s:obsetup}

We observed M32 with the red side detector of the HST/FOS for nine
spacecraft orbits on August 22, 1995 (project GO-5847). The COSTAR
optics corrected the spherical aberration of the HST primary
mirror. Telescope tracking was done in `fine lock'. The RMS telescope
jitter was typically $3$ milli-arcsec (mas). Each orbit consisted of
$54$ minutes of target visibility time, followed by $42$ minutes of
Earth occultation. The first two orbits were used to accurately
acquire the nucleus of M32. Subsequently, spectra were taken at
various positions near the nucleus. The G570H grating was used in
`quarter-stepping' mode, yielding spectra with 2064 pixels covering
the wavelength range from 4572 {\AA} to 6821 {\AA}. Periods of Earth
occultation were used to obtain wavelength calibration spectra of the
arc lamp. In the last orbit of the observations the FOS was used in a
special mode to obtain an image of the central arcsec of M32, to
verify the telescope pointing.

Two different square apertures were used to obtain the spectra: the
`0.1--PAIR' and the `0.25--PAIR'. These are the smallest apertures
available on the FOS. Although they are paired (two square openings
separated by several arcsec), they were used exclusively in
single-aperture mode. All data were collected through the upper
aperture. The aperture names are based on their size in arcsec {\it
before\/} the installation of COSTAR. Their nominal post-COSTAR sizes
are smaller: $0.086''$ square and $0.215''$ square,
respectively. Evans (1995a) presented calibration observations in
which the light throughput was measured for a point source as
function of position in the apertures. Models for these observations
presented in Appendix~\ref{s:AppA} indicate somewhat smaller aperture
sizes: $0.068''$ square and $0.191''$ square, respectively. These
sizes are adopted in the remainder of the paper.

The natural coordinate system of the FOS is the Cartesian $(X,Y)$
system defined in the FOS Instrument Handbook (Keyes \etal 1995).  The
apertures have their sides parallel to the axes of this system. The
grating disperses the light in the $X$-direction. The $(X,Y)$ system
is `left-handed' when projected onto the sky. In the following we
adopt a more convenient `right-handed' system $(x,y)$, defined through
$x=-X$ and $y=Y$. During the observations, the direction of the
$y$-axis was fixed at a position angle of $161^{\circ}$ on the
sky. This coincides with the major axis position angle of M32
(cf.~Lauer \etal 1992).

\section{Aperture positions}
\label{s:pos}

\subsection{Target acquisition}
\label{s:acq_results}

The position of the M32 center in the HST Guide Star Coordinate system
is RA $=$ 0h 42m 41.82s, $\delta = 40^{\circ} \, 51' \, 53.9''$. The
positional uncertainty is dominated by that in the HST Guide Star
Catalog itself, which is $\sim 0.5''$ RMS. Hence, some form of
acquisition is required to properly position the galaxy in the small
apertures. For an extended target with a sharp and well defined
brightness maximum, such as M32, the method of choice is a so-called
`peak-up' acquisition, which consists of a number of `stages'. Each
stage adopts a rectangular grid of $N_x \times N_y$ points on the sky,
with inter-point spacings of $s_x$ and $s_y$, respectively. An FOS
aperture is positioned at each of the grid points, and the total
number of counts is measured in some exposure time. The grid point
with the most counts is adopted as new estimate of the target
position. Subsequent stages in a sequence use smaller and smaller
apertures, with each stage increasing the accuracy of the target
positioning.

Various standard choices exist for the number of stages, and for
the aperture, the grid and the exposure time for each stage in the
acquisition (Keyes \etal 1995). These standard sequences work well for
point-source targets. However, they have not been particularly well
tested for extended sources, especially not for acquisitions into the
small 0.1-PAIR aperture. Therefore, as part of the preparation of the
observations, a software package was developed to simulate FOS peak-up
acquisitions of arbitrary targets in a Monte-Carlo manner (van der
Marel 1995). Guided by simulations done with this package, a
non-standard 5-stage peak-up sequence was constructed for M32. The
parameters of this sequence are listed in Table~\ref{t:peakup}. The
sequence was executed with the G570H grating in place, because M32 is
too bright to be acquired with the FOS mirror. The accuracy of the
sequence is such that in an idealized situation (no Poisson noise, no
telescope jitter, etc.)  $\vert \Delta_x \vert < 0.022''$ and $\vert
\Delta_y \vert < 0.022''$, where $(\Delta_x,\Delta_y)$ is the 
difference between the adopted pointing at the end of the acquisition
and the true position of the galaxy center.

\placetable{t:peakup}

The observed intensities in the acquisition stages can be analyzed
{\it post facto}, to determine the extent to which the acquisition was
successful. Most important are the results of the fifth and final
stage, in which the 0.1-PAIR aperture was sequentially positioned at
the points of a $5 \times 5$ grid on the sky, with spacing of
$0.043''$ between the grid points. The left panel of
Figure~\ref{f:peakup5} shows a grey-scale representation of the
observations. The solid dot marks the position with the highest
intensity, which was adopted by the telescope software as its best
estimate for the position of the galaxy center.

\placefigure{f:peakup5}

To model the observed intensities, the `cusp model' for the
unconvolved M32 surface brightness presented by Lauer
\etal (1992) was used, which is based on pre-COSTAR HST/WFPC images. The 
point-spread-function (PSF) of the HST+FOS was approximated by a sum
of Gaussians, as determined in Appendix~\ref{s:AppA}. The total
magnitude at each grid point was calculated using the equations in the
Appendix. The offset $(\Delta_x,\Delta_y)$ was varied to optimize the
fit. The best fit is displayed in the right panel of
Figure~\ref{f:peakup5}. It has $(\Delta_x,\Delta_y) =
(-0.010'',0.015'')$, with a formal error in either coordinate (based
on the chi-squared surface of the fit) of approximately 1 mas. The
target acquisition was thus successful.

\placefigure{f:grey_image}

\subsection{FOS image}

An FOS image of the central region of M32 was obtained in the final
orbit of the observations. The telescope was commanded to position the
upper `1.0-PAIR' aperture, which has a nominal post-COSTAR size of
$0.86''$ square, on the galaxy center, after which the intensity on
the photocathode was scanned with the diode array of the detector
(with no grating in the light path). The scan pattern was chosen to
provide $0.038'' \times 0.041''$ pixels. However, the spatial
resolution of the image is poor, $0.301'' \times 1.291''$,
corresponding to the size of one FOS diode (Koratkar \etal 1994; Evans
1995b). The resolution can be improved by deconvolution.
Figure~\ref{f:grey_image} shows the result of Lucy deconvolving the
raw image with a boxcar PSF with the size of one diode. The cross
marks the galaxy center, while the dot marks the aperture center,
i.e.,~the position where the telescope thought the galaxy center would
be. The latter is offset from the actual galaxy center by
$(\Delta_x,\Delta_y) = (-0.024'',0.095'')$ (with a formal error of 3
mas in each coordinate). This offset at the end of the observations is
much larger than that achieved by the target acquisition. Hence,
positional errors must have accumulated during the observations. The
reason for this is not well understood. However, the most likely cause
is a drift of the telescope pointing due to thermal effects related to
the heating and cooling of the spacecraft and Fine Guidance Sensors
(FGSs) during the 14.4 hours of the observations. Such thermal effects
exist (Lupie, priv.~comm.), but were not previously reported to
complicate observations with the FOS (Keyes, priv.~comm.).

\placetable{t:obsetup}

\subsection{Spectra}

Nine spectra were taken. Those obtained with the 0.1-PAIR aperture are
referred to as S1--S4, those obtained with the 0.25-PAIR aperture as
L1--L5. Table~\ref{t:obsetup} summarizes the observational setup for
each spectrum. It also lists the position with respect to the galaxy
center at which the telescope was instructed to place the aperture.
These {\it intended} aperture placements are illustrated in the left
panel of Figure~\ref{f:aper_plot}. 

\placefigure{f:aper_plot}

In reality, the apertures were not placed exactly at their intended
positions. The actual position is a sum of two vectors: (i) an
intentional offset from the (telescope's estimate of the) galaxy
center (Table~\ref{t:obsetup}); and (ii) an unintentional offset
$(\Delta_x,\Delta_y)$ of the telescope's estimate of the galaxy center
from the actual galaxy center. The intentional offsets are applied by
slewing the telescope, which should be very precise ($\lta 1$ mas
positional error). To determine the actual aperture positions one thus
needs to know the offset $(\Delta_x,\Delta_y)$ for each observation.
These can be determined by modeling the observed intensities of the
spectra, which are listed in Table~\ref{t:applaces}.

\placetable{t:applaces}

The spectra S1 and S2, taken in the subsequent orbits \#3 and \#4,
were scheduled to have identical aperture positions
(cf.~Table~\ref{t:obsetup}). The same holds for the observations L4
and L5, taken in the subsequent orbits \#8 and \#9. In both cases,
however, the observed intensities in the two spectra are significantly
different. This indicates that the aperture positions must have
changed (by $\sim 25$ mas, cf.~Figure~\ref{f:telpoint} below) between
the orbits \#3 and \#4, and between the orbits \#8 and \#9. There was
no motion of either the grating wheel or the aperture wheel of the
instrument between the different observations. The guide star
reacquisition at the beginning of each orbit is normally accurate to a
few mas. Hence, the inferred pointing drifts are indeed most likely
due to thermal effects on the FGS.

\placefigure{f:telpoint}

To proceed we make a number of simplifying assumptions about the
offset $(\Delta_x,\Delta_y)$: (i) the offset for the first spectrum,
S1, taken in orbit \#3, was identical to the offset determined from
the final peak-up stage (Figure~\ref{f:peakup5}), executed in orbit
\#2; (ii) the offset for the last spectrum, L5, taken in orbit \#9,
was identical to the offset determined from the FOS image
(Figure~\ref{f:grey_image}), taken in that same orbit; (iii) the
observations S2--S4, taken in orbits \#4, \#5 and \#6, have a common
offset; and (iv) the observations L1--L4, taken in orbits \#7 and \#8,
have a common offset. These assumptions do not allow for positional
errors between orbits \#4 and \#5, between orbits
\#5 and \#6, and between orbits \#7 and \#8. Even though errors could
have occurred, they are not required to fit the observed
intensities. The assumptions do allow for an error between orbits \#6
and \#7, when the 0.1-PAIR aperture was replaced by the 0.25-PAIR
aperture. This is because these apertures might not be exactly
concentric on the aperture wheel (Evans \etal 1995), and because the
motion of the aperture wheel might have a positional non-repeatability
(Dahlem \& Koratkar 1994).

With these assumptions, the offset $(\Delta_x,\Delta_y)$ can be
determined for each observation by fitting to the observed
intensities, using the Lauer \etal (1992) cusp model for the M32
surface brightness, and the aperture sizes and PSF determined in
Appendix~\ref{s:AppA}. The results are displayed in
Figure~\ref{f:telpoint}. The corresponding aperture placements for the
individual spectra are listed in Table~\ref{t:applaces} and are
displayed in the right panel of Figure~\ref{f:aper_plot}. The fits to
the intensities constrain only the absolute value of $\Delta_x$, and
not its sign. This introduces an ambiguity for L1--L4, which was
resolved by adopting the same sign for $\Delta_x$ as measured in
Figure~\ref{f:grey_image} (so as to minimize the size of the
positional error between orbits \#8 and \#9). At the positions of the
apertures (along the major axis), the surface brightness changes more
rapidly along the $y$ direction than along the $x$-direction. Hence,
the observed intensities constrain the $\Delta_y$ more tightly than
the $\Delta_x$. Other model assumptions to fit the observed
intensities were also studied. In all cases the results for $\Delta_y$
agreed to within a few mas, while the $\Delta_x$ could differ by as
much as $0.02''$.

\section{Data reduction}
\label{s:datareduc}

Most of the necessary data reduction steps are performed by the HST
calibration `pipeline'. For the M32 data, three issues required
additional attention: (i) wavelength calibration; (ii) absolute
sensitivity calibration; and (iii) flat-fielding.

\subsection{Wavelength calibration}

The wavelength scale provided by the calibration pipeline is not
accurate enough for our project. Arc lamp spectra were therefore
obtained in each orbit during occultation. The emission line centers
were determined, yielding for each arc spectrum $j$ and for each line
$i$ with vacuum wavelength $\lambda_{{\rm vac},i}$, the observed
wavelength $\lambda_{{\rm pipe},ij}$ on the scale provided by the
calibration pipeline. Subsequently, the observed offsets $\Delta_{ij}
\equiv \lambda_{{\rm pipe},ij} - \lambda_{{\rm vac},i}$ were fit as
\begin{equation}
\Delta_{ij} = D + P_3 (\lambda_{{\rm vac},i}) + d_j .
\end{equation}
The constant offset $D$ is the result of non-repeatability in the FOS
grating and aperture wheels, and was measured to be $6.17${\AA}. The
third order polynomial $P_3$ with zero mean accounts for a slight
($\vert P_3 \vert \lta 0.25${\AA}) non-linearity of the wavelength
scale. The offsets $d_j$ account for constant shifts of the wavelength
scale from orbit to orbit, mostly due to the `geo-magnetically induced
image motion problem' (Keyes \etal 1995). These shifts ranged between
$\pm 0.18${\AA}, and could be determined with $0.02${\AA} ($\sim 1
\kms$) accuracy.

The wavelength scale of the M32 spectra was corrected using the above
fits to the arc spectra. An additional shift of $-0.769${\AA} was
applied to correct for the fact that the light path for external
targets differs from that for the internal arc lamp (Keyes \etal
1995). For each M32 spectrum the $d_j$ were used that was/were
determined from the arc spectrum/spectra obtained immediately
preceding or following the M32 spectrum. The fact that the arc spectra
are never obtained completely simultaneously with galaxy spectra
introduces a slight additional uncertainty. The cumulative uncertainty
in the mean streaming velocities of the stars as a result of
uncertainties in the wavelength scale is estimated to be $\lta 2
\kms$. This is always smaller than the formal errors in these
velocities as a result of photon noise (cf.~Table~\ref{t:kinresults}
below).

After wavelength calibration (and absolute sensitivity calibration and
flat-fielding as described below) the galaxy spectra were rebinned
logarithmically in wavelength, as required for the stellar kinematical
analysis. A scale of $57.9 \kms /$pixel was used. The FOS is a photon
counting detector, so the formal errors in the spectra follow directly
from Poisson statistics. On average, the (logarithmic) rebinning
decreases the formal errors, since it reduces the scatter between
neighboring pixels. The decrease can be calculated for each pixel, and
was taken into account in the subsequent analysis. The correlation
between the errors in neighboring pixels induced by the rebinning was
neglected.

\subsection{Absolute sensitivity calibration}

The calibration pipeline multiplies the observed count rate spectra by
a so-called `inverse sensitivity file' (IVS), to obtain fluxes in
${\rm erg} \, {\rm cm}^{-2} \, {\rm s}^{-1} \,${\AA}$^{-1}$. The
pipeline IVS files contain a wavelength dependent aperture throughput
correction, based on calibration observations of point sources. These,
however, are not applicable to extended sources. Therefore, instead of
the standard pipeline IVS file a more recent IVS file with no aperture
throughput correction was used, provided to us by FOS Instrument
Scientist Tony Keyes.

Each wavelength in the M32 spectra samples a slightly different region
of the galaxy, because of the wavelength dependence of the PSF. This
influences only the continuum slope of the spectra, because the PSF
varies very slowly with wavelength. In the stellar kinematical
analysis one is only interested in the absorption lines, which are not
influenced. The underlying continuum is subtracted. Hence, no attempt
was made to construct aperture throughput corrections. (In fact, this
would require accurate knowledge of the wavelength dependence of the
PSF, which is not readily available for the FOS; see
Appendix~\ref{s:AppA}).

\subsection{Flat-fielding}

A G570H flat-field based on multiple 0.25-PAIR (upper) aperture
observations of a star was used, as provided to us by Tony Keyes. No
flat-fields obtained explicitly with the 0.1-PAIR (upper) aperture
were available. A star illuminates the photocathode of the detector
differently than an extended source. To test the appropriateness of
the flat-field for the M32 data it was cross-correlated with the
continuum subtracted normalized galaxy spectra. This yielded clear
peaks, indicating that the flat-field and the M32 spectra share the
same features (as they should).  The peak was only $\sim $10\% lower
for the 0.1-PAIR spectra than for the 0.25-PAIR spectra. This
indicates that the flat-field can be properly used for the 0.1-PAIR
(upper) aperture as well, although a flat-field obtained specifically
with that aperture would have been preferable. Shifts between the
galaxy spectra and flat-field were identified from the positions of
the cross-correlation peaks, and were corrected for (see also Kormendy
\etal 1996a). The flat-fielding removes most of the pixel-to-pixel
sensitivity variations, but some small residual variations might
remain.

\section{Line-spread-functions}
\label{s:lsf}

The observed spectrum of a source is the convolution of its actual
spectrum with the line-spread-function ${\rm LSF}(\lambda)$. The LSF
can be written as the convolution of an `illumination function'
$B(\lambda)$ and an `instrumental broadening function' $H(\lambda)$
(cf.~Appendix~\ref{s:AppB}). The function $B(\lambda)$ is the
normalized intensity distribution of the light that falls onto the
grating. It depends on the target brightness and choice of
aperture. The direction of dispersion is parallel to the $x$
direction. Hence, $B(\lambda)$ is the integral over the $y$ direction
of the two-dimensional brightness distribution that falls onto the
grating. For the PSF and aperture sizes derived in
Appendix~\ref{s:AppA}, $B(\lambda)$ is given by
equations~(\ref{Blamdef}) and~(\ref{Bkernel}) in
Appendix~\ref{s:AppB}. It can be calculated for each observation upon
substitution of the Lauer \etal (1992) cusp model for the M32 surface
brightness, and the aperture positions in Table~\ref{t:applaces}. The
normalized function $H(\lambda)$ accounts for the instrumental
broadening due to the grating and finite size of a detector diode (the
resolution element). It was determined empirically from fits to the
emission line shapes in the arc spectra, as discussed in
Appendix~\ref{s:AppB}.

The LSF has zero mean if the galaxy light is distributed symmetrically
within the aperture. This is not generally the case, because the
aperture centers are not exactly at $x=0$ (Figure~\ref{f:aper_plot}).
Calculations show that the intensity weighted mean position is within
$0.01''$ from the aperture center for all the observations.
Nonetheless, there are noticeable wavelength shifts, because $0.01''$
projects onto $0.145${\AA} in the wavelength direction ($8.4 \kms$ at
5170{\AA}). Table~\ref{t:applaces} lists the LSF mean as calculated
for each of the M32 spectra, both in {\AA} and in $\kms$ at
5170{\AA}. Errors in these values due to errors in the aperture
positions do not exceed 0.04{\AA} ($2 \kms$ at 5170{\AA}).

\placefigure{f:lsf}

The width and shape of the LSF are determined mainly by the aperture
size. The calculated LSFs for all the M32 spectra are shown in
Figure~\ref{f:lsf}. The LSF shapes for the 0.1-PAIR observations are
very similar to each other, as are the LSF shapes for the 0.25-PAIR
observations. The shapes are also similar to the emission line shapes
observed directly in arc spectra (Figure~\ref{f:arc}). The 0.25-PAIR
LSF can be well approximated by convolving the 0.1-PAIR LSF by a
Gaussian with a dispersion of $1.14${\AA} ($66 \kms$ at 5170{\AA}), as
illustrated by the dotted curve in Figure~\ref{f:lsf}.

The best-fitting Gaussian to the 0.1-PAIR LSF has a dispersion of
$1.65${\AA} ($96 \kms$ at 5170{\AA}). For the 0.25-PAIR LSF it has a
dispersion of $1.98${\AA} ($115 \kms$ at 5170{\AA}). However, these
numbers are of limited use in characterizing the LSFs, since these
have noticeably broader wings than a Gaussian.

\section{Template spectrum}
\label{s:temp}

For stellar kinematical analysis a template spectrum is needed to
compare the M32 spectra to. To avoid systematic errors it is important
to minimize template mismatching. This is best done by constructing a
composite template which contains an appropriate mix of spectral
types. Observing template stars with the HST is inefficient and time
consuming (mainly because of the lengthy target acquisitions that are
required). To date less than a handful different template stars have
been observed with the HST, all of similar spectral type. These HST
spectra are insufficient to construct an optimal template.  A
ground-based template library of spectra of 27 stars of different
spectral types was therefore used, obtained in February 1990 by
M.~Franx at the 4m telescope of the KPNO with the RC Spectrograph (as
discussed previously by van der Marel \& Franx 1993). The LSF of these
spectra is approximately Gaussian with a dispersion of $1.22${\AA}
($71 \kms$ at 5170{\AA}). The spectra cover the spectral range from
4836{\AA} to 5547{\AA}, centered on the Mg b triplet at $\sim
5170${\AA}. This is the most useful wavelength range for stellar
kinematical analysis, so it is no drawback that this range is smaller
than the full range covered by the FOS G570H grating.

The stellar spectra were similarly rebinned logarithmically as the
galaxy spectra, and shifted to a common velocity. No attempt was made
to construct a different composite template for each M32 spectrum
(which is reasonable, given that there are no strong color gradients
in the central arcsec; Lugger \etal 1992). Instead, the best fit was
sought to a grand-total M32 spectrum, constructed by summing the HST
spectra. The resulting composite template is a weighted mix of the
individual templates, with the weights determined using the method
outlined in Appendix~A.3 of van der Marel (1994a). The mix contains
giants, sub-giants and dwarfs of spectral types~G and~K.

The composite template is used in the remainder of the paper. However,
other templates were studied as well. For example, a stellar
kinematical analysis was performed with a K-star template spectrum
obtained from the HST Data Archive, taken with the FOS circular 0.3
aperture ($0.26''$ diameter) by H.~Ford and collaborators before the
installation of COSTAR. This template provides a poor fit to the
spectrum of M32. Nonetheless, the stellar kinematical properties
inferred with this template were found to be consistent with those
derived using the composite template.

\section{Kinematical analysis}
\label{s:kinematics}

\subsection{Description}

The stellar kinematical analysis was performed with the method
presented in van der Marel (1994a). It fits the convolution of a
parametrized velocity profile and a template spectrum to a galaxy
spectrum in pixel space, using chi-squared minimization. The formal
errors in the fit parameters follow from the shape of the chi-squared
surface near its minimum. The method has been well tested, and its
results agree with those from other methods for extracting stellar
kinematics from galaxy spectra. Deviations of the line-of-sight
velocity profiles from Gaussians contain useful information on the
dynamical structure of galaxies. Unfortunately, the velocity
resolution of the FOS is too poor to extract any reliable velocity
profile shape information from the M32 data. The analysis was
therefore restricted to Gaussian velocity profile fits.

Data obtained with the FOS is time resolved. The red side detector
reads and stores the data every $\sim 2$ minutes. Between 6 and 20
spectra were therefore available per aperture position.  The aperture
positions for the observations L4 and L5 differ by $\sim 0.024''$, but
both are at a distance of $\sim 0.5''$ from the galaxy center, where
kinematical gradients are small. In the kinematical analysis L4 and L5
were therefore treated as a single observation, referred to as L45.
Especially for the 0.1-PAIR observations, the signal-to-noise ratio
(S/N) of the individual read-outs is low. Individual read-outs for a
given aperture position were therefore added to achieve an average S/N
of $\sim 10$ per (logarithmically rebinned) pixel. This resulted in 4
independent spectra for each of the observations S1--S4 with the
0.1-PAIR aperture, and between 9 and 11 independent spectra for the
observations L1, L2, L3 and L45 with the 0.25-PAIR aperture. A stellar
kinematical analysis was performed on each of these spectra. The
kinematical results for each aperture position were then averaged
together, weighted with the errors. It was found that this yields
slightly more accurate results than a kinematical analysis of the sum
of the individual spectra for a given aperture position. No systematic
trends with time in the orbit were found in the kinematical results.

The stellar kinematical analysis was performed over the wavelength
range 4859--5520{\AA}. Regions influenced by so-called `noisy' diodes
were masked. The fit to the galaxy spectrum includes a low order
polynomial to account for continuum differences between the galaxy and
template spectrum. This polynomial is fit simultaneously with the
velocity profile. The parameters of the best-fitting velocity profile
are virtually independent of the choice of polynomial order. A
polynomial of order 5 was used.

\subsection{Corrections}
\label{s:correc}

The mean velocity and velocity dispersion of the best-fitting Gaussian
velocity profiles must be corrected for instrumental effects. The mean
velocity is biased because the LSF mean is non-zero. This bias is
removed by subtracting the corrections $\Delta V_{\rm LSF}$ listed in
Table~\ref{t:applaces}. These were calculated for the wavelength
5170{\AA} of the Mg b triplet, but the wavelength dependence of the
velocity corrections over the fit range is negligible ($\lta 0.4 \kms$
in absolute value). Further, to obtain velocities in the M32 frame one
must subtract a constant offset, determined by the difference between
the systemic velocity of M32 and the template velocity. These were not
known accurately enough to determine the offset directly, and it was
therefore estimated from the data itself. Somewhat arbitrarily, it was
fixed to the intercept of the best least-squares fit line in a plot
(see Figure~\ref{f:allres}c below) of rotation versus position $y$ along
the M32 major axis for the observations S4, S1 and S2 (for which
$y=-0.046''$, $0.015''$ and $0.040''$, respectively).

The velocity dispersions $\tsigma$ of the best-fitting Gaussians must
be corrected for differences in the LSFs of the template and galaxy
spectra. For this it is useful to consider not the observed
dispersions themselves, but rather differences in the observed
dispersions between two observations.  Let $\tsigma_{\rm A}$ and
$\tsigma_{\rm B}$ be the best-fitting dispersions for observations at
positions A and B. Assume that the LSF of observation B can be
obtained from the LSF of observation A through convolution with a
Gaussian of dispersion $S_{\rm LSF}$. The difference between the
actual dispersions at positions A and B can then be estimated as:
\begin{equation}
   \sigma_{\rm A}^2 - \sigma_{\rm B}^2 = \tsigma_{\rm A}^2 -
   \tsigma_{\rm B}^2 + S_{\rm LSF}^2 ,
\label{sigdef}
\end{equation}
(which uses the fact that the convolution of two Gaussians is again a
Gaussian, the dispersion of which is the RMS sum of the input
dispersions). This approach has two important advantages. First, the
LSF of the template influences the kinematical analysis at points A
and B in the same way, so it does not enter into $\sigma_{\rm A}^2 -
\sigma_{\rm B}^2$. Second, the differences in the LSFs for the HST observations
can in fact to good approximation be accounted for by convolution with
a Gaussian (cf.~Figure~\ref{f:lsf}). One does {\it not\/} have to
assume that the LSFs themselves are Gaussian, which would be a poor
approximation.

In the following, the observation L45 at $\sim 0.5''$ from the galaxy
center is used as `reference point'. Equation~(\ref{sigdef}) then
yields
\begin{equation}
   \sigma^2 - \sigma_{\rm L45}^2 =
          \tsigma^2 - \tsigma_{\rm L45}^2 + S_{\rm LSF}^2 ,
\label{sigcor}
\end{equation}
where $S_{\rm LSF}=0$ for an observation with the 0.25-PAIR aperture,
and $S_{\rm LSF} = 66.2 \kms$ for an observation with the 0.1-PAIR
(based on the results of Section~\ref{s:lsf}). The latter value was
calculated for the wavelength 5170{\AA} of the Mg b triplet, which is
roughly at the center of the fit range.  The actual value of $S_{\rm
LSF}$ varies from $62.0$ to $70.4 \kms$ over the fit range, which can
be taken into account by assigning a `formal' error of $4.2 \kms$ to
$S_{\rm LSF}$. Equation~(\ref{sigcor}) can be used to estimate the
stellar velocity dispersion $\sigma$ for any of the observations
S1--S4 and L1--L3, if one assumes that the dispersion $\sigma_{\rm
L45}$ at the position of observation L45 is known a priori. At $\sim
0.5''$ along the major axis one is beyond the region most influenced
by seeing in ground-based data. Based on the data of van der Marel
\etal (1994a) and Bender, Kormendy \& Dehnen (1996) it was assumed that
$\sigma_{\rm L45} = 70 \pm 5 \kms$ (see Figure~\ref{f:allres}d below).

It is demonstrated in Appendix~\ref{s:AppC} that the velocity
dispersions thus obtained have no systematic biases, and that even
velocity dispersions as small as $\sim 50 \kms$ can be measured
reliably, in spite of the large FOS instrumental dispersion of $\gta
100 \kms$. The formal errors in dispersions $\sigma$ obtained from
equation~(\ref{sigcor}) follow from the errors in $\tsigma$,
$\tsigma_{\rm L45}$, $\sigma_{\rm L45}$ and $S_{\rm LSF}^2$ using a
standard first order error analysis. Systematic errors due to template
mismatching are not likely to exceed a few $\kms$, and are smaller
than the formal errors in the measurements.

\placetable{t:kinresults}

\subsection{Results}
\label{ss:results}

Table~\ref{t:kinresults} lists the line strengths, rotation velocities
and velocity dispersions, after correction for all instrumental
effects. These results are discussed below. Figure~\ref{f:allres}
displays the results as function of the position $y$ of the aperture
center along the M32 major axis. The best spatial resolution
ground-based M32 data are also shown. These are the data from van der
Marel \etal (1994a) obtained at the William Herschel Telescope (WHT),
and those of Bender, Kormendy \& Dehnen (1996) obtained at the Canada
France Hawaii Telescope (CFHT). A comparison of these ground-based
data to older (lower spatial resolution) ground-based data can be
found in Kormendy \& Richstone (1995).

The spatial resolution for each of the observations in
Figure~\ref{f:allres} is characterized by a convolution kernel, which
results from the PSF and finite aperture size (for ground-based
long-slit data the effective aperture is the spatial CCD pixel size by
the slit width). The top left-panel illustrates the difference in
spatial resolution between the data sets. It displays the projection
of the convolution kernel along the M32 minor axis, i.e., the
probability $P(y)$ that a photon observed in a given aperture was
emitted at distance $y$ from the center of that aperture, measured
along the M32 major axis. These probability distributions have FWHM
values of $0.102''$, $0.203''$, $0.541''$ and $0.911''$, for the
HST/FOS/0.1-PAIR, HST/FOS/0.25-PAIR, CFHT and WHT data, respectively.
The HST data therefore provide a factor 5 increase in spatial
resolution over even the best available ground-based data.

\placefigure{f:allres}

\subsubsection{Line strengths}
\label{s:gamma}

The line strengths shown in Figure~\ref{f:allres}b were normalized to
an average of unity (line strengths are measured with respect to the
template spectrum, and are hence on an arbitrary scale). There is no
obvious trend with radius. The observations with the 0.1-PAIR aperture
closest to the nucleus have relatively low line strength, but this is
unlikely to be real. Metallicities and line strengths generally
increase towards the centers of galaxies. The observed low values
could be an artifact due to the presence of a black hole. The stars
that move rapidly near the black hole produce broad velocity profile
wings. These can be mistaken for an enhanced continuum, which causes
the line strength to be underestimated (van der Marel 1994b).  The
broad-band colors of M32 are constant in the central arcsec (Lugger
\etal 1992). The true line strengths in the central $0.5''$ covered by
the HST observations are therefore probably also close to constant.

The observed line strengths have a sizable RMS scatter of $\sim
0.055$. This exceeds the formal errors, which are on average $\sim
0.03$. The measurements are therefore not statistically consistent
with a constant line strength (as is confirmed by a chi-squared
test). The scatter in the observed line strengths could be the result
of minor inaccuracies in the data, such as unidentified bad pixels or
flat-field errors, or it could be real, and due to shot noise in the
contributions from individual stars. The total V-band luminosity
inside the aperture varies from $3 \times 10^4 L_{\odot}$ (for
observation S3) to $2.5 \times 10^5 L_{\odot}$ (for observation
L3). Most of the observed light comes from giants, some of which can
be as luminous as $10^3 L_{\odot}$ (Freedman 1989). The luminosity of
a `typical' giant as measured from surface brightness fluctuations
(Tonry, Ajhar \& Luppino 1990) is $\sim 10^{1.6} L_{\odot}$. This
implies that there are $N = 10^{3-4}$ typical stars in each HST
aperture. Therefore, $\sqrt{N}$ fluctuations can be of order a few
percent of the intrinsic line strength differences between stars.

\subsubsection{Kinematics} 

Figure~\ref{f:allres}c shows the rotation velocities. The HST rotation
curve is smooth. It is significantly steeper in the central $0.2''$
than seen in the ground-based data, consistent with the presence of a
central black hole. The rotation velocity is observed to be $\sim 30
\kms$ at $0.1''$ from the nucleus. This is roughly twice the value
measured from the ground at this distance.

The scatter in the line strength measurements (Figure~\ref{f:allres}b)
induces scatter in the velocity dispersion measurements, because the
velocity dispersion and line strength of a Gaussian fit are
statistically correlated. Tests indicate that a line strength error
$\Delta \gamma$ induces an error $\Delta \sigma = 1.4 \, (\Delta
\gamma/0.01) \kms$ in the velocity dispersion. Rotation velocity measurements
are not influenced. To avoid unwanted errors in the velocity
dispersion measurements, they were determined by fitting Gaussian
velocity profiles to the data while keeping the line strength constant
to the mean of the measurements in Figure~\ref{f:allres}b. This yields
velocity dispersions that differ only at the $\lta 10 \kms$ level from
those obtained by Gaussian fits with variable line strength, so the
main results do not depend on this correction.

Figure~\ref{f:allres}d shows the resulting velocity dispersions. The
dispersion measured with the 0.1-PAIR aperture closest to the nucleus
is $156 \pm 10 \kms$. The profile of the velocity dispersion with
radius is smooth, with the exception of the measurement $\sigma = 96
\pm 10 \kms$ from observation S2, which is significantly lower than
that for the other observations in the central $0.1''$. We doubt the
reality of this, but have not been able to attribute it to any known
instrumental effect. The average of the four dispersion measurements
inside $0.1''$ is $126 \kms$, with a RMS scatter of $21 \kms$. Hence,
the central velocity dispersion is definitely higher than the
85--95$\kms$ determined from ground-based measurements. It thus
appears that a Keplerian increase in the velocity dispersion close to
the center of M32 has been resolved.

The difference in the dispersions measured for observations~S1 and~S2,
which were obtained with the same aperture at similar positions,
indicates that uncertainties in the velocity dispersions are dominated
by systematic errors, rather than Poisson noise. Systematic errors in
the rotation velocity measurements are believed to be much smaller,
primarily because these are less sensitive to the poor instrumental
resolution of the FOS. Either way, the data clearly imply that the
dispersion in the central $0.1''$ is higher than measured from
ground-based data.  Whether the nuclear dispersion seen through a
small aperture is indeed as high as $156 \kms$ should await
verification by observations with the future HST spectrograph STIS.

\section{Discussion}
\label{s:discussion}

HST/FOS spectra were obtained through small apertures centered on the
nuclear region and major axis of M32. Kinematical analysis of these
data is more complicated than it is for ground-based data: the FOS is
a low-resolution spectrograph, and is not particularly well suited for
dynamical studies of low-dispersion objects such as M32.  One of the
main results of this paper is therefore that stellar kinematical
quantities {\it can} in fact be determined from the data, provided
that a detailed study is made of various technical issues, including:
calibration and modeling of the FOS aperture sizes, PSF and LSFs,
determination of the aperture positions from the data, accurate
wavelength calibration, template matching, and corrections for
instrumental resolution.

The resulting kinematical quantities are displayed in
Figure~\ref{f:allres}. The main result is that in the central $0.1''$
the M32 rotation curve is significantly steeper, and the velocity
dispersion significantly higher, than measured from even the highest
resolution ground-based data. These observational results are
robust. They do not depend critically on the corrections made in the
analysis; even a rough analysis without any template matching and
corrections for instrumental effects shows these main features. The
results are exactly what would be expected if M32 does indeed have a
massive dark object in its nucleus, as suggested previously on the
basis of ground-based data. In the companion papers in this series
(van der Marel \etal 1997a,b; Cretton \etal 1997) we construct dynamical
models to address the presence of such a dark object quantitatively.
We also address the question whether the dark object has to be a black
hole, or whether plausible alternatives still exist.


\acknowledgments

The authors are grateful to the FOS Instrument Scientists Tony Keyes,
Anuradha Koratkar and Michael Dahlem for helpful discussions, and to
Bill Workman and Jean Surdej for successful scheduling and
implementation of the observations. Tony Keyes provided preliminary
instrument calibration data and Marijn Franx provided the template
library. Simon White helped with the early stages of the
project. Support for this work was provided by NASA through grant
number \#GO-05847.01-94A, and through a Hubble Fellowship
\#HF-1065.01-94A awarded to RPvdM, both from the Space Telescope
Science Institute which is operated by the Association of Universities
for Research in Astronomy, Incorporated, under NASA contract
NAS5-26555.


\clearpage

\appendix

\section{Point-spread-function and aperture sizes}
\label{s:AppA}

Some calibrations exist of the HST+FOS PSFs and LSFs (Koratkar 1996)
and of the FOS aperture sizes (Evans \etal 1995), but these are too
crude for the purposes of this paper. To obtain more accurate
calibrations we present detailed models for the calibration
observations obtained by Evans (1995a) after the installation of
COSTAR. He centered a star in FOS apertures, and stepped the star
across each aperture in the FOS $x$ and $y$ directions,
respectively. At each step the total intensity was measured. Here the
results are considered for the FOS red detector with three apertures:
the square upper aperture of the 0.1-PAIR, the square upper aperture
of the 0.25-PAIR, and the circular 1.0. The names of these apertures
are based on their size in arcsec, {\it before} the installation of
COSTAR. Their nominal post-COSTAR sizes are smaller by a factor $\sim
0.86$.  Figure~\ref{f:transmis} shows for each aperture the observed
intensity as function of distance from the aperture center, normalized
by the observed intensity with the star at the center of the circular
1.0 aperture. The aperture length or diameter is easily determined
from these data for apertures much larger than the PSF core: it is
then twice the radius at which the intensity has fallen to 50\% of its
central value. This was used by Evans \etal (1995) to estimate the
size of all FOS apertures. Although this is reasonable for the larger
apertures, it is not for the smaller apertures. For those a more
careful analysis is required, in which the aperture sizes and PSF are
fitted to the data simultaneously.

For a target with surface brightness $S(x,y)$, the intensity $I$
observed through an aperture centered on $(x,y)$ is
\begin{equation}
   I (x,y) = \int_{-\infty}^{\infty} \int_{-\infty}^{\infty}
     S(x',y') \> K(x'-x,y'-y) \> dx' \> dy' ,
\label{Itot}
\end{equation}
where $K(x,y)$ is a convolution kernel that depends on the PSF and the
aperture size and geometry. The PSF is assumed to be a circularly
symmetric sum of Gaussians,
\begin{equation}
   {\rm PSF}(r) = \sum_{i=1}^N  \>
     {{\gamma_i}\over{2\pi \sigma_i^2}} \> 
     \exp \left [- \fr{1}{2} ( \fr{r}{\sigma_i} )^2 \right ] ,
\label{PSFGauss}
\end{equation}
where the $\gamma_i$ must satisfy $\sum_{i=1}^N {\gamma_i} = 1$. For
a perfectly rectangular aperture with sides $A_x$ and $A_y$ the
convolution kernel is then
\begin{equation}
  K (x,y) = 
     \sum_{i=1}^N \> {{\gamma_i}\over 4} \>
     \left \lbrace \mathop{\rm erf}\nolimits
            \left [ \fr{x+(A_x/2)}{\sqrt{2}\> \sigma_i} \right ]
           -\mathop{\rm erf}\nolimits
            \left [ \fr{x-(A_x/2)}{\sqrt{2}\> \sigma_i} \right ]
     \right \rbrace \>
     \left \lbrace \mathop{\rm erf}\nolimits
            \left [ \fr{y+(A_y/2)}{\sqrt{2}\> \sigma_i} \right ]
           -\mathop{\rm erf}\nolimits
            \left [ \fr{y-(A_y/2)}{\sqrt{2}\> \sigma_i} \right ]
     \right \rbrace ,
\end{equation}
while for a perfectly circular aperture with diameter $D$ it is 
\begin{equation}
  K (\rho \cos \theta,\rho \sin \theta) =
       \sum_{i=1}^N \> {{\gamma_i}\over {\sigma_i^2}}
       \exp \left [- \fr{1}{2} ( \fr{\rho}{\sigma_i} )^2 \right ] 
    \int_0^{D/2} 
       \exp \left [- \fr{1}{2} ( \fr{r}{\sigma_i} )^2 \right ] \>
      I_0(\fr{r \rho}{\sigma_i^2}) \> r \> dr ,
\end{equation}
(van der Marel 1995). The function ${\rm erf}(t)$ is the error
function and $I_0(t)$ is a Bessel function.

\placefigure{f:transmis}

This model was used to fit the data in Figure~\ref{f:transmis}. For a
point source at $(x_0,y_0)$, the intensity observed through an
aperture centered at the origin is simply $K(x_0,y_0)$. The solid
curves in the figure show the best chi-squared fit to the data for a
PSF that is a sum of three Gaussians. The aperture sizes were treated
as free parameters. The square apertures were assumed to have equal
sizes in the $x$ and $y$ directions. The model was forced to fit 2
additional constraints, obtained from theoretical modeling (as quoted
in Keyes \etal 1995): (i) the throughput (for a centered point source)
of the circular 1.0 aperture is $0.96$ times that of a rectangular
aperture of size $3.7'' \times 1.3''$; and (ii) the absolute
throughput of the $3.7'' \times 1.3''$ aperture is $0.975$. The
best-fitting model parameters are listed in Table~\ref{t:psfapfit}.
The fit is robust, in spite of the fact that the model parameters are
highly correlated. In particular: (i) no good fit can be obtained if
the aperture sizes are held fixed to their nominal post-COSTAR values
(which are $0.086''$ square, $0.215''$ square, and $0.86''$ circular,
respectively); and (ii) the fit does not improve significantly if the
number of Gaussians in the PSF is increased. The formal errors in the
best fit parameters are $0.01$ for the $\gamma_i$, and $2$\% for the
$\sigma_i$ and aperture sizes. These numbers are not too meaningful,
however, because systematic errors probably exceed the formal errors.

\placetable{t:psfapfit}

\placefigure{f:psffit}

Figure~\ref{f:psffit} shows the PSF for the best fit, and the
encircled flux curve $E(r)$. The latter is given by
\begin{equation}
   E(r) \> \equiv \> \int_0^r {\rm PSF}(r') \> 2\pi r' \> dr' 
        \> = \> 1 - \sum_{i=1}^N {\gamma_i} \>
                \exp \left [- \fr{1}{2} ( \fr{r}{\sigma_i} )^2 \right ] .
\label{encircled}
\end{equation}
The PSF has 73\% of the light within a circle of radius $0.1''$. This
is smaller than the 84\% measured from star images at visual
wavelengths obtained with the FOC, but larger than the 65\% measured
with the WFPC2 (which suffers from additional pixel scattering). It
should be noted that the PSF derived here pertains only to the upper
position of the square paired FOS apertures. The aperture
transmissions, and thus the PSF, are somewhat different at the lower
positions of the square paired apertures (Evans 1995a).

The above analysis ignores the effects of diffraction at the aperture
edges. The derived aperture sizes are therefore not the geometrical
apertures sizes. This might explain why the derived sizes are smaller
than the nominal post-COSTAR sizes. Non-circularly symmetric features
in the PSF are also ignored. Such features might in fact be present,
given that the results of scanning a star across the aperture in the
$x$ and $y$ directions differ slightly (Figure~\ref{f:transmis}).
Either way, both the aperture sizes and the PSF enter into the data
analysis only through the convolution kernels $K(x,y)$. These kernels
are adequately fit by the model, independent of whether or not the
aperture sizes and PSF of the model are unbiased estimates of their
true values.

The observations in Figure~\ref{f:transmis} were derived from
so-called `white light' images, i.e., with no grating in the light
path. The PSF with the parameters of Table~\ref{t:psfapfit} is
therefore an average over all wavelengths to which the detector is
sensitive. The M32 observations were obtained with the G570H
grating. Observations with this grating (Bohlin \& Colina 1995) show
fractional throughputs for centered point sources that are larger by a
few percent than those in Figure~\ref{f:transmis}. This indicates that
the core of the PSF for the G570H grating might be $\sim 15$\% smaller
than that of the PSF derived here. On the other hand, in fitting the
M32 peak-up acquisition observations (Section~\ref{s:acq_results}) by
convolving the Lauer \etal (1992) cusp model for the M32 surface
brightness, it was found that the chi-squared of the fit could be
improved by {\it increasing} the width of the PSF core by $\sim
15$\%. This could mean that the PSF does in fact have a broader core
than derived here, or alternatively, that the Lauer \etal model, based
on WFPC1 measurements, is not a perfectly adequate representation of
the M32 surface brightness in the central $0.1''$. In the absence of
theoretical constraints on the wavelength dependence of the PSF from
accurate wavefront modeling, it was decided not to make any
corrections to the PSF derived from the white light images
(Table~\ref{t:psfapfit}). The PSF remains somewhat uncertain, but it
was verified that changes in the PSF core width between $-20$\% and
20\% do not change any of the major conclusions of our paper(s).

\section{Line-spread-function}
\label{s:AppB}

The observed spectrum $O(\lambda)$ of a source is the convolution of
its actual spectrum $S(\lambda)$ with the
line-spread-function ${\rm LSF}(\lambda)$,
\begin{equation}
   O(\lambda) = \int_{-\infty}^{\infty} d \lambda' \>
        S(\lambda') \> {\rm LSF}(\lambda - \lambda') .
\end{equation}
The LSF can be written as the convolution
\begin{equation}
   {\rm LSF}(\lambda) = \int_{-\infty}^{\infty}
        d \lambda' \> B(\lambda') \> H(\lambda-\lambda')   ,
\label{LSFdef}
\end{equation}
where the `illumination function' $B(\lambda)$ is the normalized
intensity distribution of the light that falls onto the grating, and
the normalized `instrumental broadening function' $H(\lambda)$
accounts for the broadening due to the grating and the detector.

The $(x,y)$ coordinate system is defined such that the $x$ direction
is parallel to the direction of dispersion. The function $B(\lambda)$
is therefore proportional to the integral over the $y$ direction of
the two-dimensional brightness distribution that falls onto the
grating. All the light in the aperture is detected, since the
apertures used for the observations are smaller than the $y$-size of
the diode array of the detector ($1.29''$). For the case of
observations of a target with surface brightness $S(x,y)$ through an
aperture centered on position $(x_0,y_0)$,
\begin{equation}
   B(\lambda) = 
      {1\over{I(x_0,y_0)}} \int_{-\infty}^{\infty} \int_{-\infty}^{\infty}
         S(x',y') \> {\cal K}(x'-x_0, y'-y_0; x) \> dx' \> dy' ,
\label{Blamdef}
\end{equation}
where $I(x_0,y_0)$ is defined in equation~(\ref{Itot}), and ${\cal
K}(x',y';x)$ is a convolution kernel that depends on the PSF and the
aperture size and geometry. For the G570H grating the wavelength
$\lambda$ is related to the position $x$ according to $\lambda =
14.489\>$(x/arcsec) {\AA}. If the aperture is rectangular of size
$A_x \times A_y$, and the PSF is as given in
equation~(\ref{PSFGauss}), then
\begin{equation}
   {\cal K}(x',y';x) = \left\{ \begin{array}{ll}
     \sum_{i=1}^N \> 
       {{\gamma_i}\over{\sqrt{8\pi}\>\sigma_i}} \> 
       \exp[-{1\over 2}({{x'-x}\over{\sigma_i}})^2] \>
       \left \lbrace \mathop{\rm erf}\nolimits
              \left [ {{y'+(A_y/2)} \over {\sqrt{2}\> \sigma_i}} \right ]
             -\mathop{\rm erf}\nolimits
              \left [ {{y'-(A_y/2)} \over {\sqrt{2}\> \sigma_i}} \right ]
       \right \rbrace ,
          & \quad \vert x \vert \leq {{A_x}\over2}        ; \\
     0 ,  & \quad \vert x \vert > {{A_x}\over2}           . 
                               \end{array} \right.
\label{Bkernel}
\end{equation}

The grating disperses the image of the aperture onto the photocathode
of the detector, which is scanned using an array of diodes. A simple
model for the instrumental broadening function $H(\lambda)$ is
therefore:
\begin{equation}
   H (\lambda) = \int_{-\infty}^{\infty}
        d \lambda' \> D(\lambda') \> G(\lambda-\lambda')   ,
\end{equation}
where $D(\lambda)$ is the normalized response function of a detector
diode and $G(\lambda)$ is the normalized broadening function of the
grating. Let $D(\lambda)$ be a top-hat function with full width $d_x$,
and let the grating broadening function be $G(\lambda) = (1/w)
f(\lambda/w)$, where $f$ is a normalized function, and $w$ is a free
parameter that measures the width of $G$. The notation
\begin{equation}
  F^{(1)} (t) = \int f(t) \> dt , \qquad
  F^{(2)} (t) = \int F^{(1)} (t) \> dt ,
\end{equation}
is used for the primitives of $f$. The function $H$ is then
\begin{equation}
  H(\lambda) = \fr{1}{d_x} \sum_{i=1}^2 (-1)^{i} F^{(1)} (t_i) , \qquad
  t_i   \equiv \fr{1}{w} [ \lambda + (-1)^{i} \fr{d_x}{2} ] .
\end{equation}
It can be determined empirically by fitting to the emission lines in
the arc spectra. If the arc lamp illuminates the aperture
homogeneously, then $B(\lambda)$ is a normalized top-hat function of
full width $A_x$. For an emission line that is intrinsically a
delta-function at $\lambda_0$, the observed line shape will be
\begin{equation}
  O_{\rm arc}(\lambda) = \fr{w}{A_x d_x} \sum_{i=1}^2 \sum_{j=1}^2
     (-1)^{i+j} F^{(2)} (t_{ij}) , 
     \qquad
  t_{ij}  \equiv \fr{1}{w} [ \lambda - \lambda_0 + 
                 (-1)^{i} \fr{A_x}{2} + (-1)^{j} \fr{d_x}{2} ] .
\label{Oarc}
\end{equation}
Several functional forms were considered for $f$. A Gaussian did not
yield good fits to the observed line shapes. Satisfactory results were
obtained with a Lorentzian, for which
\begin{equation}
  f(t)       = \fr{1}{\pi (1+t^2)}    , \qquad
  F^{(1)}(t) = \fr{1}{\pi} \arctan t  , \qquad
  F^{(2)}(t) = \fr{1}{\pi} [ t \arctan t - \fr{1}{2} \ln (1+t^2) ] .
\end{equation}
With this choice, the function $H$ has two parameters, $w$ and $d_x$.
The best fit is generally attained for values of $d_x$ that differ
from the actual size of one diode, which is $4.36${\AA} for the G570H
grating. This probably indicates that the true diode response function
is not a perfect top-hat. However, this does not invalidate the model.
One is interested only in a fit to the function $H(\lambda)$, not in a
determination of either $D(\lambda)$ or $G(\lambda)$. Therefore, both
$w$ and $d_x$ are treated as free fitting parameters.

\placefigure{f:arc}

The best fits were determined to the emission line shapes in arc
spectra obtained with the 0.1-PAIR and the 0.25-PAIR apertures, while
keeping the values of $A_x$ fixed to their values in
Table~\ref{t:psfapfit}. Figure~\ref{f:arc} shows the fits to the arc
lines and the best-fitting function $H$. No evidence was found for a
wavelength dependence of the LSF over the wavelength range covered by
the grating.  The best-fit parameters are: $w = 0.595 \pm 0.02${\AA}
and $d_x = 3.807 \pm 0.02${\AA} for the 0.1-PAIR aperture, and $w =
0.816 \pm 0.04${\AA} and $d_x = 3.904 \pm 0.04${\AA} for the 0.25-PAIR
aperture. Hence, the function $H(\lambda)$ is broader for the
0.25-PAIR aperture than for the 0.1-PAIR. The difference is too large
to be attributed to small errors in the assumed aperture sizes. Hence,
the instrumental broadening is more complicated in reality than in our
simplified model, in which one would have expected $H(\lambda)$ to be
independent of the aperture size. However, the only important point
for the interpretation of the M32 spectra is that an accurate
empirical description of the LSF is available. The fits in
Figure~\ref{f:arc} suggest that the model is satisfactory for this
purpose.

\section{Velocity dispersion tests}
\label{s:AppC}

\subsection{Simulated galaxy spectra}
\label{s:Ctests}

Tests with artificial galaxy spectra were performed to determine
whether systematic biases could be present in the velocity dispersion
measurements. Artificial galaxy spectra were created from each of the
27 stellar spectra in the template library, by broadening them with
Gaussian velocity profiles with dispersions $\sigma_{\rm in}$. Twenty
copies were made of each broadened template, and noise was added to
yield a S/N of 10 per pixel. A stellar kinematical analysis was then
performed on each copy, using the composite spectrum discussed in
Section~\ref{s:temp} as template. The twenty velocity dispersion
measurements where then averaged to yield an output velocity
dispersion $\sigma_{\rm out}$.

\placefigure{f:kintest}

Figure~\ref{f:kintest} plots $\sigma_{\rm out}$ as function of
$\sigma_{\rm in}$ for each of the input spectra. The results lie
closely along the line $\sigma_{\rm out} = \sigma_{\rm in}$. The minor
deviations can be attributed entirely to the template mismatch in the
simulations (the input stellar spectra are giants, sub-giants and
dwarfs of spectral types G, K and M). At dispersions below $\sim 50
\kms$ there is some tendency for the dispersion to be
underestimated. However, for the M32 observations one is always in the
situation where the difference in dispersion between the galaxy and
template spectrum is $\gta 100 \kms$ (because the LSF of the galaxy
spectra is broader than that of the template, in addition to the
kinematical Doppler broadening). At these dispersions there are no
significant biases in the results of the kinematical analysis, but
only small ($\lta 5 \kms$) errors due to template mismatching. These
errors are nearly independent of the velocity dispersion. Since the
approach used in this paper for determining velocity dispersions
relies on differences between velocity dispersion measurements at
different positions, it is expected that even these small errors
largely cancel out.

\subsection{Consistency check}

A crude consistency check on our approach for determining velocity
dispersions can be obtained by making the (poor) assumption that the
LSFs of both the galaxy and template spectra are Gaussian, with
dispersions $\sigma_{\rm LSF,t}$ and $\sigma_{\rm LSF,g}$,
respectively. The stellar velocity dispersion $\sigma$ then follows
from the dispersion ${\tsigma}$ obtained by fitting to the galaxy
spectrum, according to:
\begin{equation}
  \sigma^2 = \tsigma^2 + \sigma_{\rm LSF,t}^2 - \sigma_{\rm LSF,g}^2 .
\label{sigcorgau}
\end{equation}
For the observation L45 the dispersion measured directly from the
spectra is $\tsigma = 111 \pm 4 \kms$. Upon substitution in
equation~(\ref{sigcorgau}) of $\sigma_{\rm LSF,t} = 71 \kms$
(cf.~Section~\ref{s:temp}), and $\sigma_{\rm LSF,g} = 115 \kms$
(cf.~Section~\ref{s:lsf}) one obtains the estimate $\sigma = 63 \pm 7
\kms$ for the stellar velocity dispersion at $\sim 0.5''$ along the M32
major axis. This agrees reasonably well with the value of $\sigma = 70
\pm 5 \kms$ adopted in Section~\ref{s:correc} on the basis of ground-based
measurements. This confirms the result of Appendix~\ref{s:Ctests} that
there are no systematic biases in the velocity dispersion measurements
from the spectra, and that the differences in the LSFs of the template
and galaxy spectra can be fully corrected for.

\clearpage


\clearpage


\def\figcapone{The final peak-up acquisition stage positions the
0.1-PAIR aperture at the points of a $5 \times 5$ grid on the sky. The
left panel shows the observed intensities in a grey-scale
representation. The dot marks the grid point with the highest
intensity, which is adopted by the telescope software as its estimate
for the position of the galaxy center. The major axis of the galaxy
lies along the $y$-axis. The right panel shows a model fit to the
observations, based on the Lauer \etal (1992) cusp model for the M32
surface brightness, and the PSF and aperture size derived in
Appendix~{\refsAppA}. The cross marks the position of the galaxy
center in this best-fit model. The adopted position for the galaxy
center is offset from the model position by (only)
$(\Delta_x,\Delta_y) = (-0.010'',0.015'')$. Hence, the target
acquisition was successful.\label{f:peakup5}}

\def\figcaptwo{FOS image with overlaid contours of the central region
of M32, taken at the end of the observations to verify the telescope
pointing.  Raw FOS images have very poor spatial resolution ($0.301''
\times 1.291''$, i.e.~similar in size to the image itself). The
displayed image was obtained from the raw image through
deconvolution. Features such as the deviations of the isophotes from
ellipses are artifacts of this deconvolution, and should not be
trusted. However, the galaxy center can be accurately determined
(cross). The dot marks the position where the telescope thought the
galaxy center would be, which is offset by $(\Delta_x,\Delta_y) =
(-0.024'',0.095'')$ from the actual center. This indicates that
pointing errors must have accumulated during the observations.
\label{f:grey_image}}

\def\figcapthree{The left panel shows the intended aperture positions
for the M32 spectra, as listed in Table~{\reftobsetup}. The aperture
sizes used in this panel are the nominal post-COSTAR sizes. The
ellipses schematically represent the isophotes of the galaxy. The
right panel shows the actual aperture positions during the
observations, determined as described in the text, and listed in
Table~{\reftapplaces}. The aperture sizes used in this panel are the
more accurate sizes determined in Appendix~{\refsAppA} from
calibration observations of a star.\label{f:aper_plot}}

\def\figcapfour{Offsets between the telescope's estimate of the galaxy
center (dots) and the true position of the galaxy center (cross),
determined as described in the text. Each offset is labeled by the
spectra to which it pertains. The offsets increased systematically
during the course of the observations.
\label{f:telpoint}}

\def\figcapfive{Four solid and five dotted curves (mostly overlaying
each other) display the LSFs calculated for the M32 spectra. The LSFs
are not centered exactly on wavelength zero, because the galaxy light
is not distributed symmetrically within the aperture. The LSF shapes
depend mainly on the aperture size. A dashed curve shows the
convolution of the 0.1-PAIR LSF with a Gaussian with a dispersion of
$1.14${\AA}. The result provides a good fit to the 0.25-PAIR LSF.
\label{f:lsf}}

\def\figcapsix{Main results of the HST/FOS observations, compared to
the ground-based data of van der Marel \etal (1994a) obtained at the
WHT, and that of Bender, Kormendy \& Dehnen (1996) obtained at the
CFHT. A legend to the symbols and line types is given in
panel~[a]. For clarity, panel~[d] includes the labels for the HST
observations that are used in the tables. The abscissa in all panels
is the position $y$ along the M32 major axis. Panel~[a] illustrates
the spatial resolution of each of the observations. The function
$P(y)$ is the probability that a photon observed in a given aperture
was emitted at a major axis distance $y$ from the center of that
aperture. The spatial resolution of the HST data is superior to that
of the ground-based data. Panel~[b] shows the line strengths derived
from the HST observations, normalized to unity. There is significant
scatter, but no trend with radius. Panels~[c] and~[d] show the
rotation velocities and velocity dispersions. The WHT data are
connected by a line for illustration. The error bars of these data are
smaller than the plot symbols. The HST rotation curve in the central
$\sim 0.2''$ is significantly steeper than that of the ground-based
data. The average dispersion in the central $0.1''$ clearly exceeds
the ground-based measurements. These kinematical properties are
consistent with the presence of a nuclear black hole in M32.
\label{f:allres}}

\def\figcapseven{Aperture transmissions for the square 0.1-PAIR upper
aperture, the square 0.25-PAIR upper aperture, and the circular 1.0
aperture, as measured by Evans (1995a). The abscissa $r$ is the
distance of a star from the aperture center. Open triangles indicate
offsets along the FOS $x$-axis, open circles indicate offsets along
the FOS $y$-axis.  The plotted transmission is defined as the ratio of
the observed intensity to the intensity observed when the star is
centered in the 1.0 aperture. The curves are the model predictions for
the PSF and the aperture sizes described in the text and
Table~{\reftpsfapfit}.
\label{f:transmis}}

\def\figcapeight{The PSF and encircled energy $E$ for the best-fit
model to the data in Figure~{\refftransmis}, the parameters of which
are listed in Table~{\reftpsfapfit}.\label{f:psffit}}

\def\figcapnine{The left panel shows the superposed data points for 12
different emission lines at various wavelengths in arc spectra
obtained with the 0.1-PAIR (circles) and 0.25-PAIR (crosses)
apertures. The curves show fits to these data for the LSF models
described in the text. There are minor discrepancies between the
predictions and the data, but overall the fits are satisfactory. The
right panel shows the normalized instrumental broadening function
$H(\lambda)$ that enters into the LSF as described in the text. This
function is broader for the 0.25-PAIR aperture (dotted curve) than for
the 0.1-PAIR aperture (solid curve). The horizontal bars indicate the
aperture sizes when projected onto the wavelength
direction.\label{f:arc}}

\def\figcapten{Results of tests of the kinematical analysis. Average
output velocity dispersions are plotted for artificial galaxy spectra,
created by convolving spectra of stars of various spectral types with
Gaussian velocity profiles with dispersions $\sigma_{\rm in}$ ranging
from 0 to $200 \kms$, and adding noise. Apart from small systematic
errors due to template mismatching, the results closely follow the
line $\sigma_{\rm out} = \sigma_{\rm in}$. In the context of the M32
observations only the simulations with $\sigma_{\rm in} \gta 100 \kms$
are relevant, as discussed in the text.\label{f:kintest}}


\ifsubmode
\figcaption{\figcapone}
\figcaption{\figcaptwo}
\figcaption{\figcapthree}
\figcaption{\figcapfour}
\figcaption{\figcapfive}
\figcaption{\figcapsix}
\figcaption{\figcapseven}
\figcaption{\figcapeight}
\figcaption{\figcapnine}
\figcaption{\figcapten}
\clearpage
\else\printfigtrue\fi

\ifprintfig


\begin{figure}
\centerline{\epsfbox{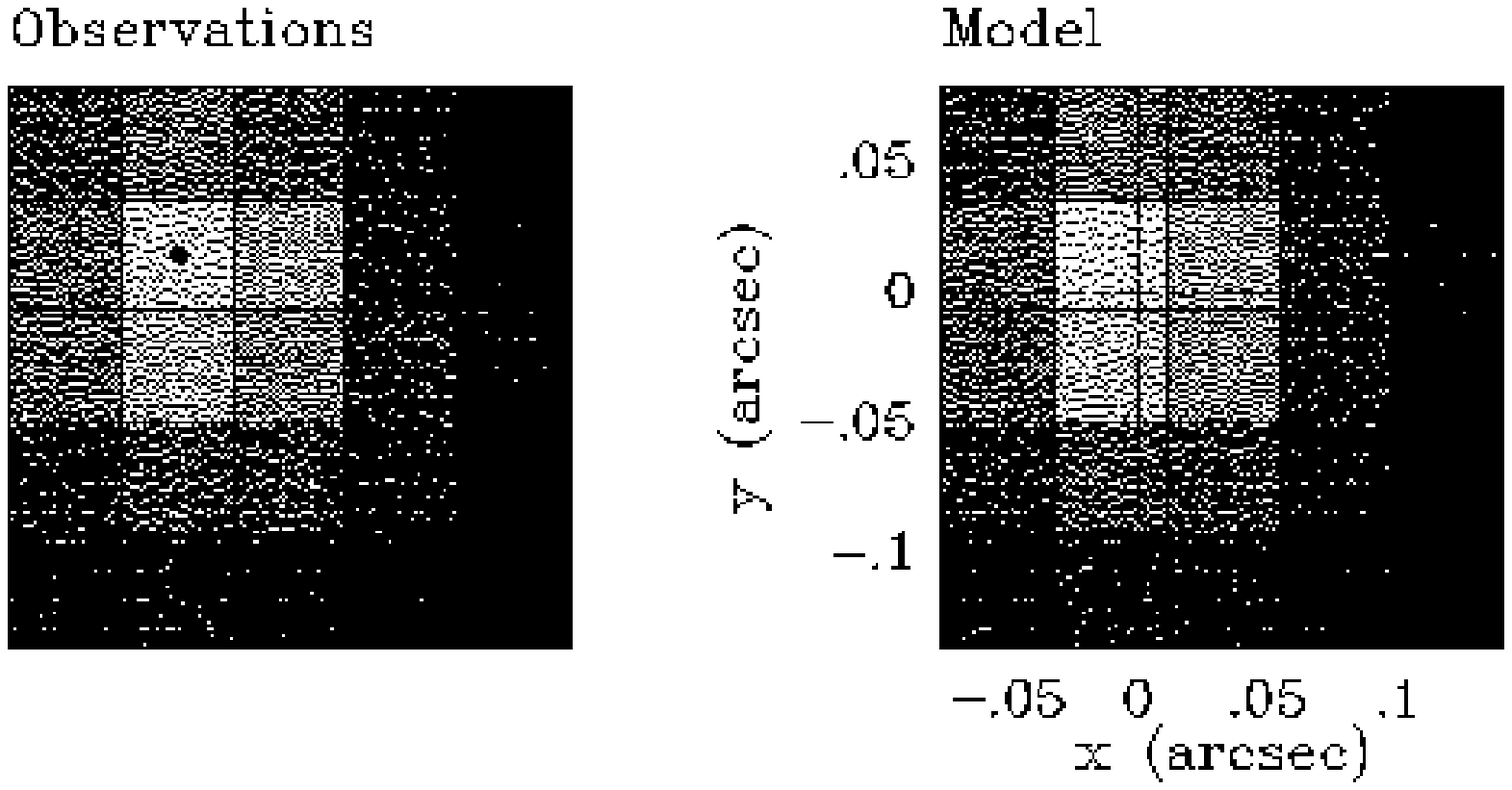}}
\ifsubmode
\vskip3.0truecm
\centerline{Figure~1}\clearpage
\else\figcaption{\figcapone}\fi
\end{figure}


\begin{figure}
\centerline{\epsfbox{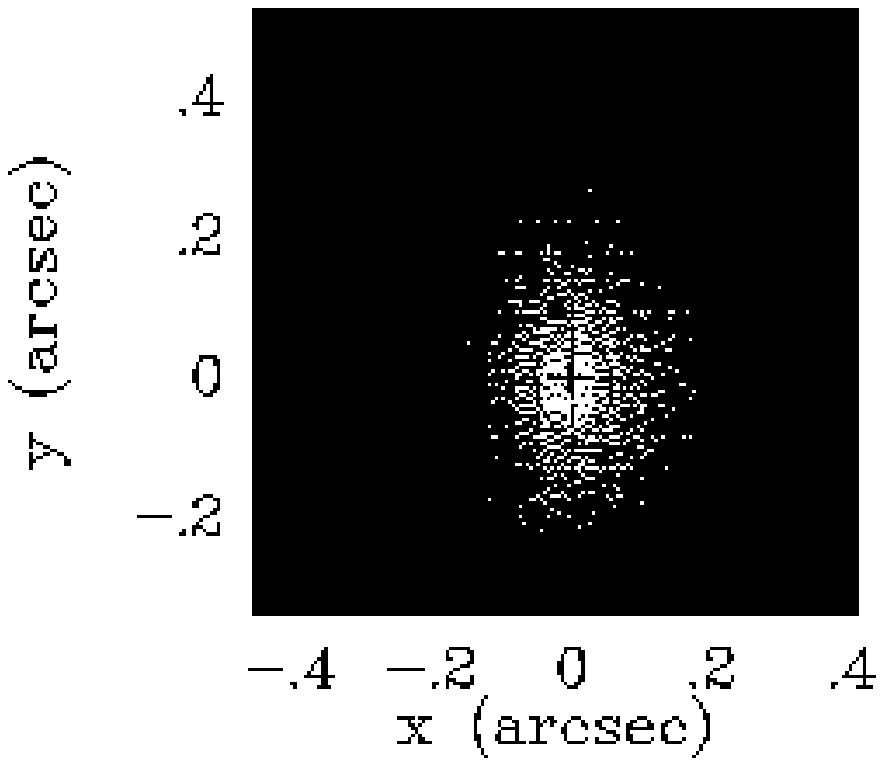}}
\ifsubmode
\vskip3.0truecm
\centerline{Figure~2}\clearpage
\else\figcaption{\figcaptwo}\fi
\end{figure}


\begin{figure}
\epsfxsize=16.0truecm
\centerline{\epsfbox{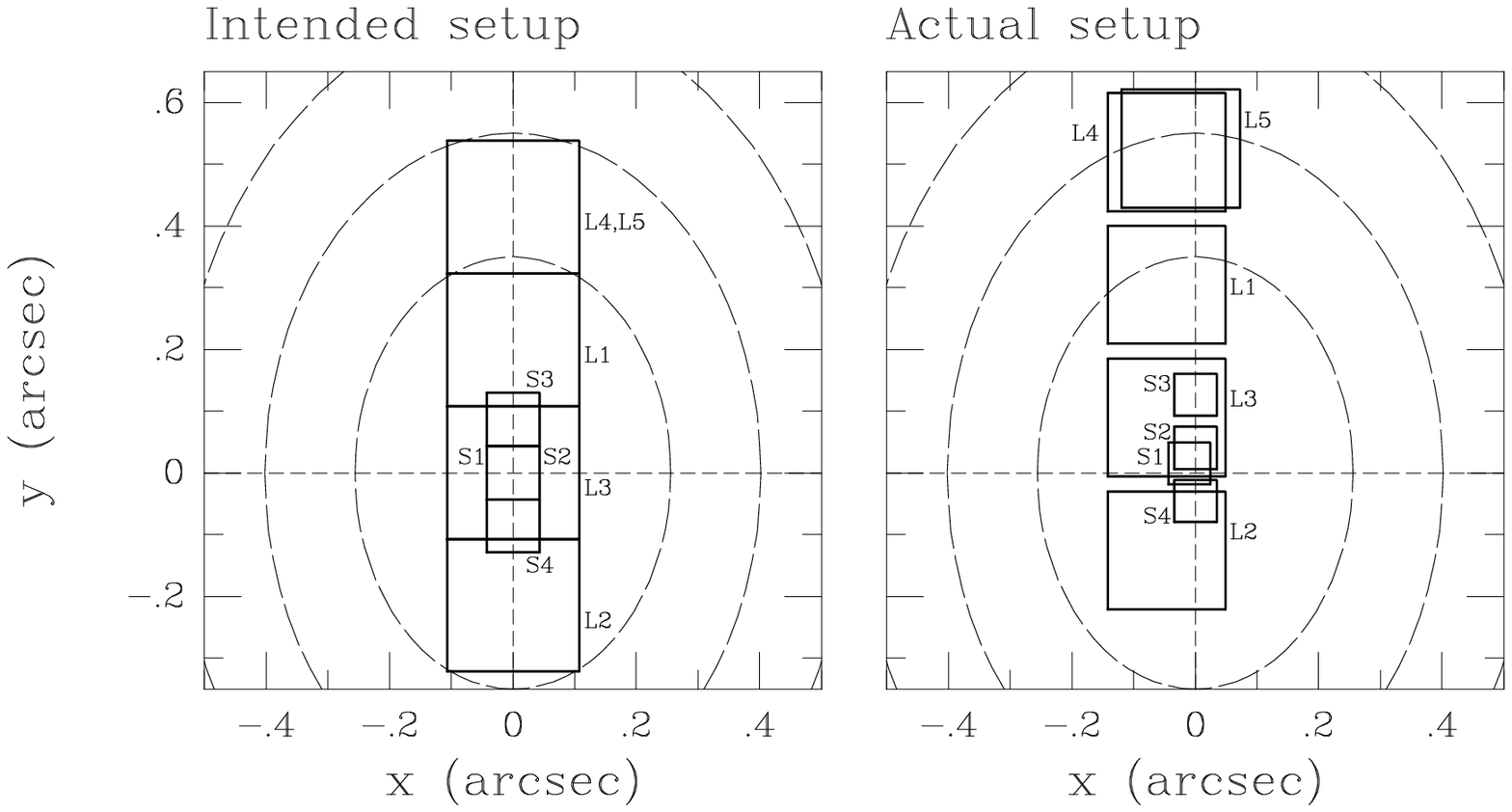}}
\ifsubmode
\vskip3.0truecm
\centerline{Figure~3}\clearpage
\else\figcaption{\figcapthree}\fi
\end{figure}


\begin{figure}
\centerline{\epsfbox{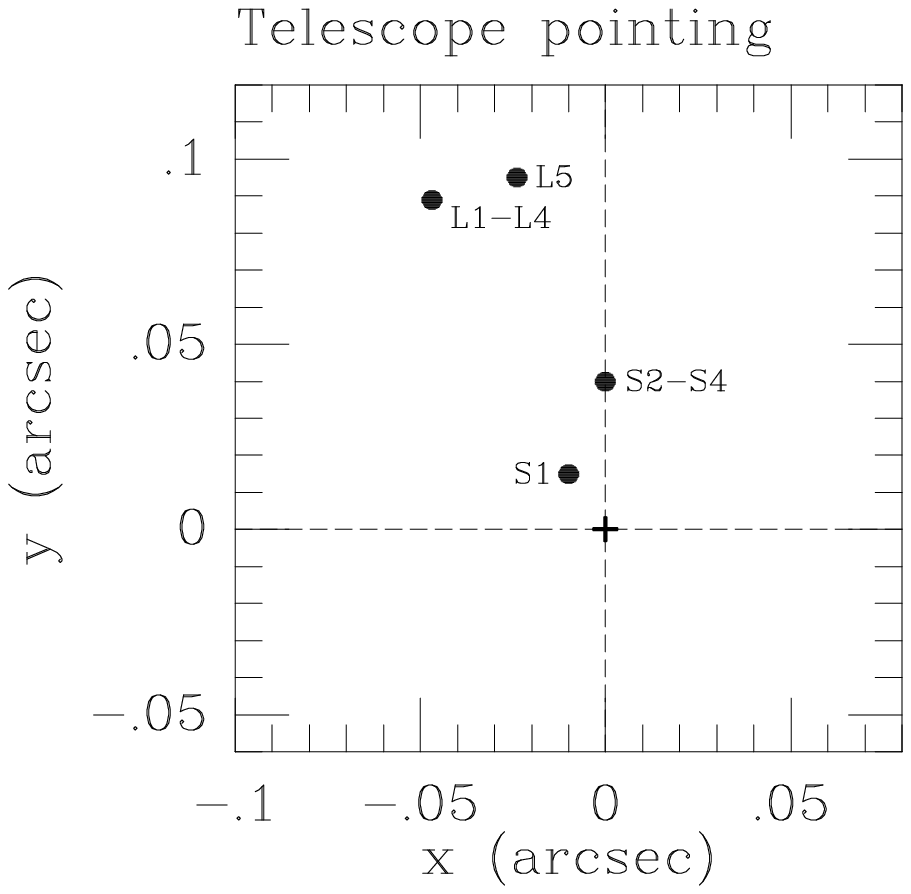}}
\ifsubmode
\vskip3.0truecm
\centerline{Figure~4}\clearpage
\else\figcaption{\figcapfour}\fi
\end{figure}


\begin{figure}
\centerline{\epsfbox{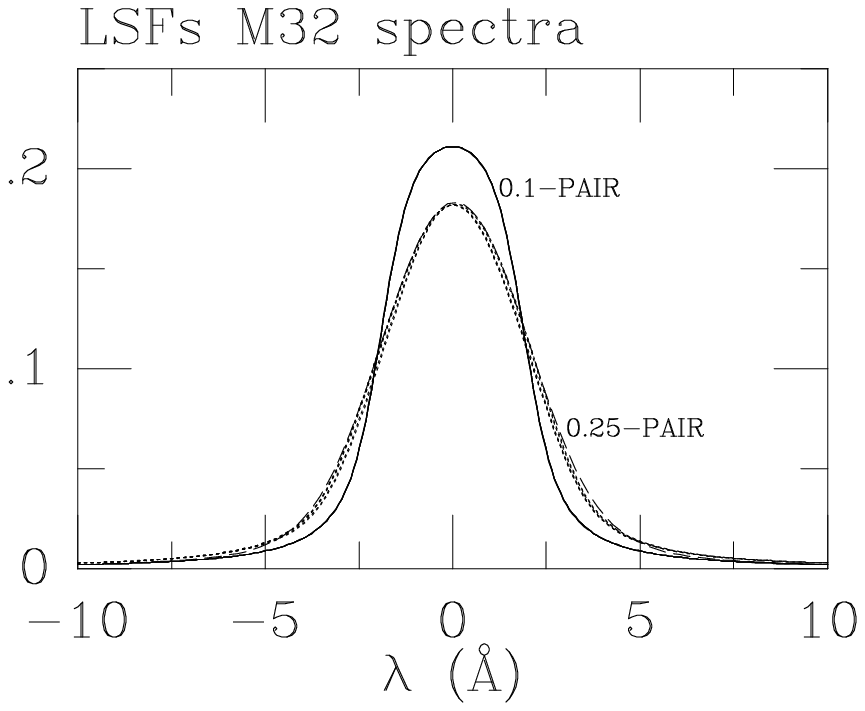}}
\ifsubmode
\vskip3.0truecm
\centerline{Figure~5}\clearpage
\else\figcaption{\figcapfive}\fi
\end{figure}


\begin{figure}
\epsfxsize=16.5truecm
\centerline{\epsfbox{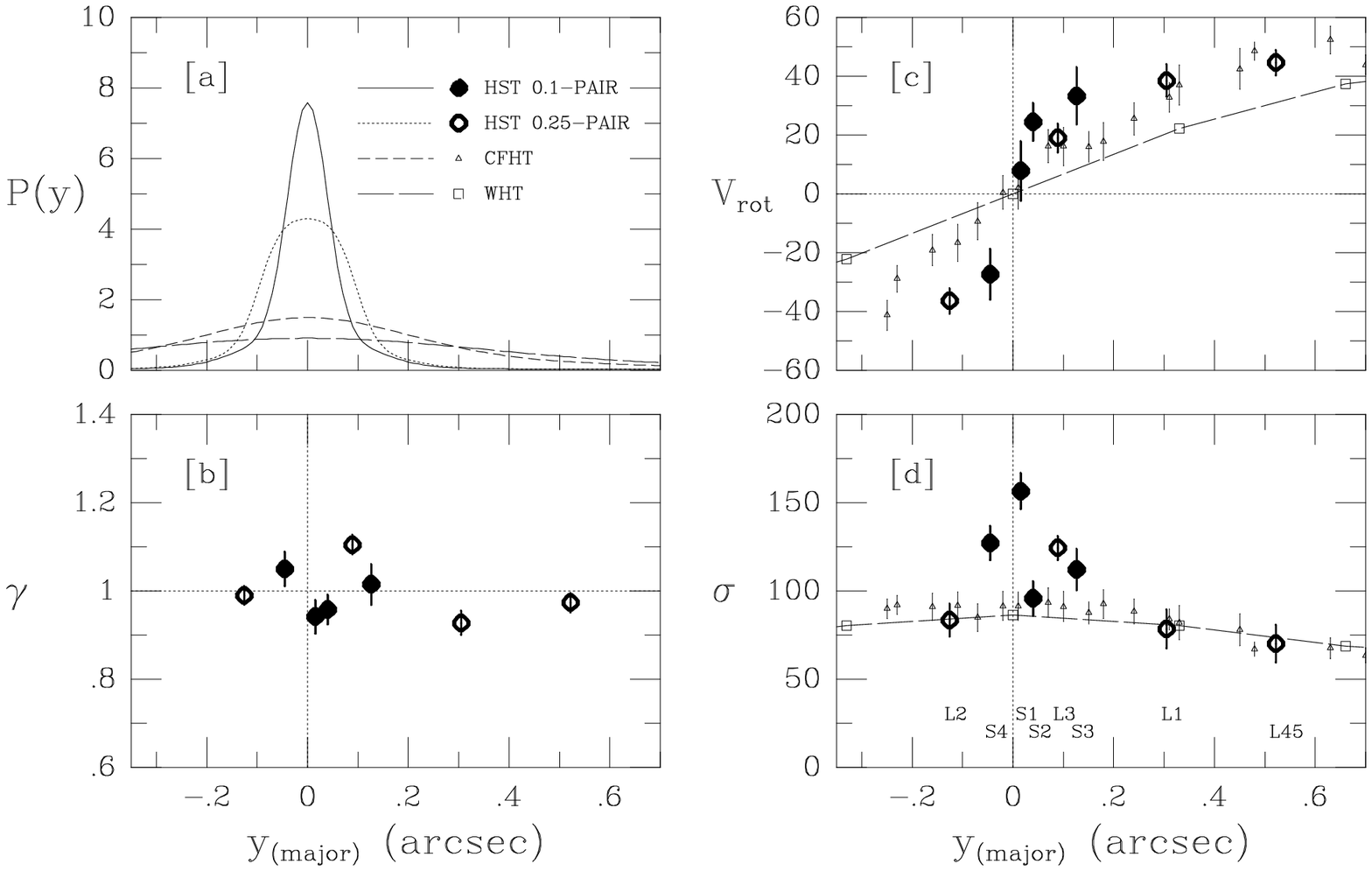}}
\ifsubmode
\vskip3.0truecm
\centerline{Figure~6}\clearpage
\else\figcaption{\figcapsix}\fi
\end{figure}


\begin{figure}
\centerline{\epsfbox{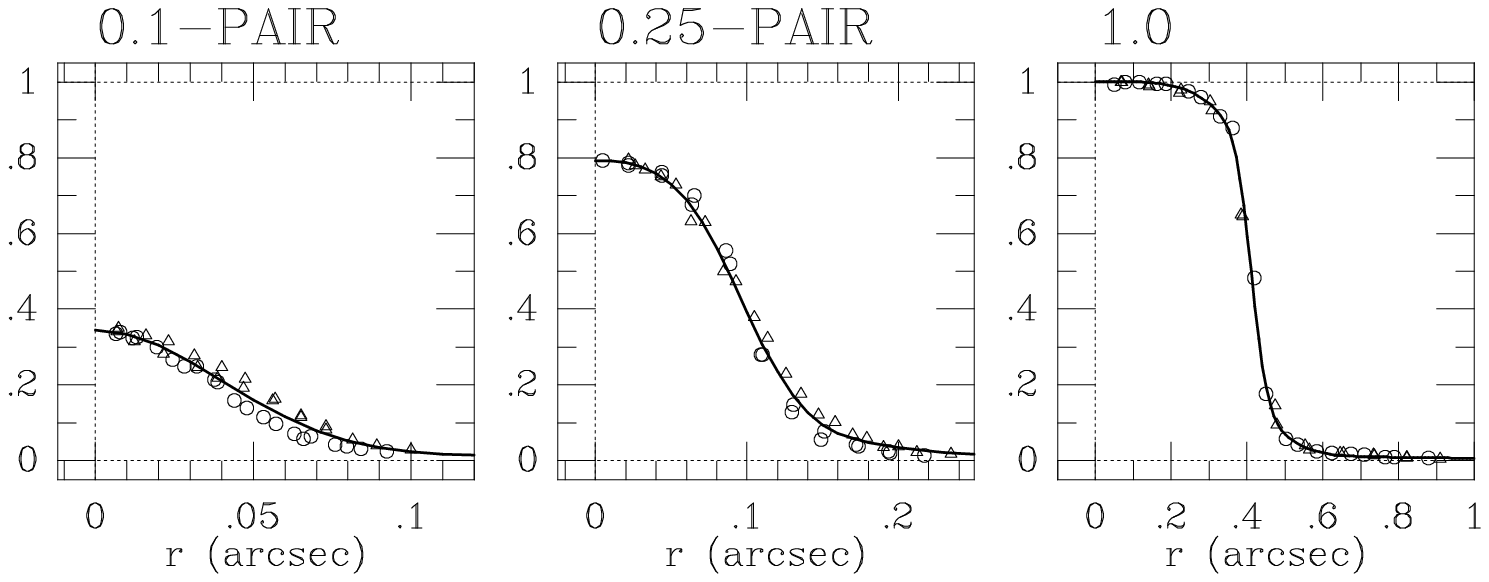}}
\ifsubmode
\vskip3.0truecm
\centerline{Figure~7}\clearpage
\else\figcaption{\figcapseven}\fi
\end{figure}


\begin{figure}
\centerline{\epsfbox{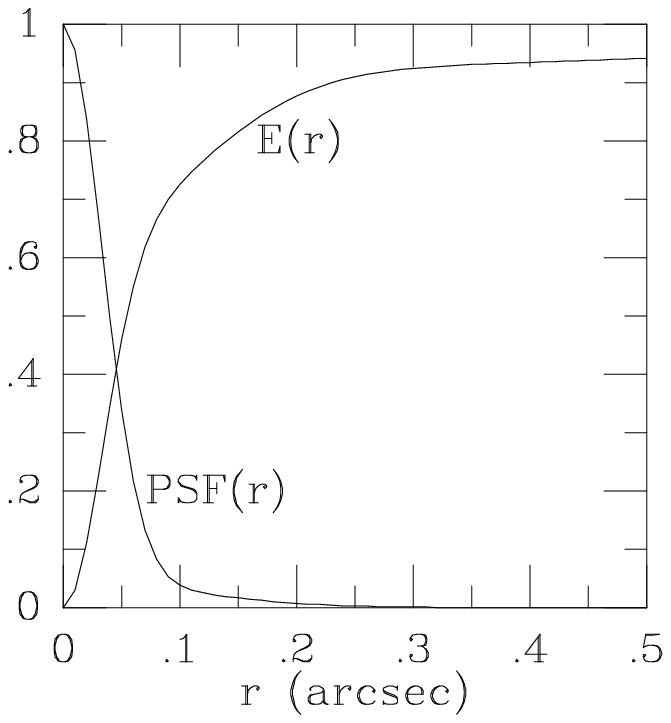}}
\ifsubmode
\vskip3.0truecm
\centerline{Figure~8}\clearpage
\else\figcaption{\figcapeight}\fi
\end{figure}


\begin{figure}
\centerline{\epsfbox{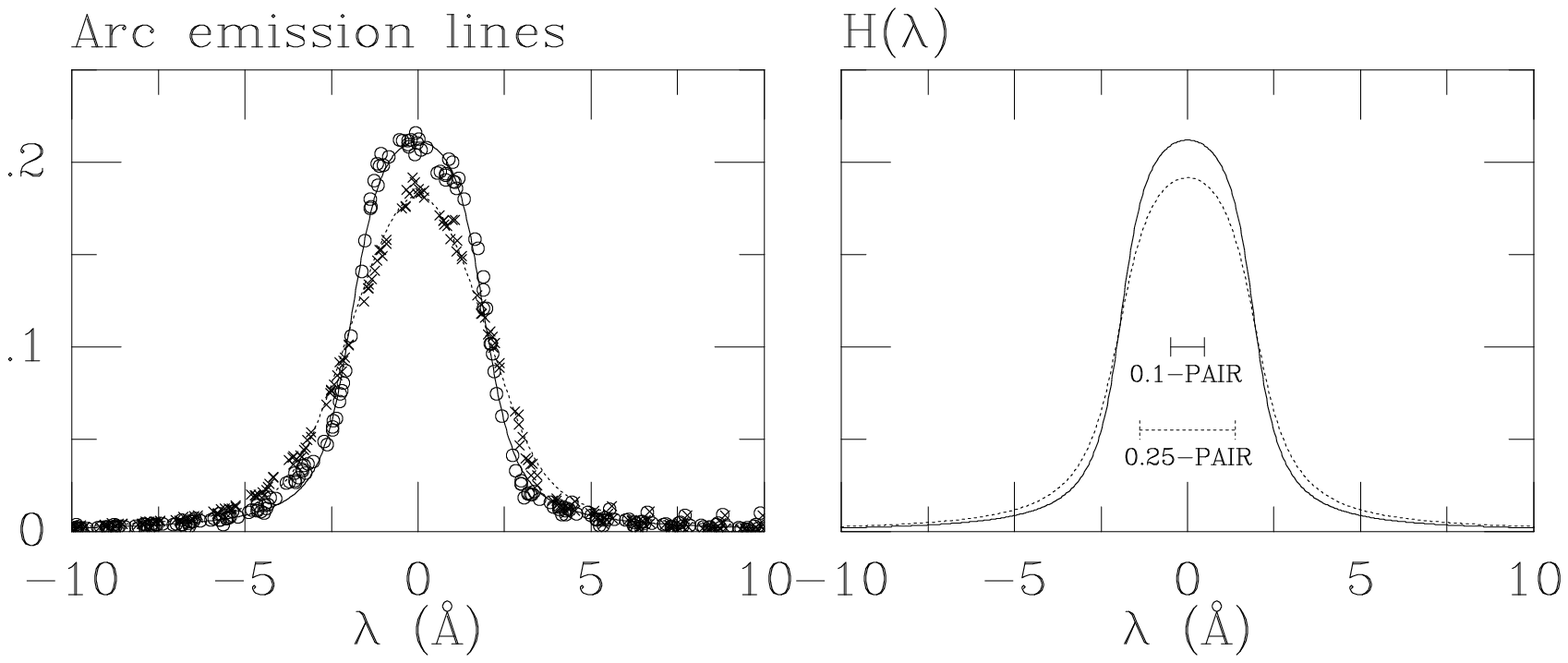}}
\ifsubmode
\vskip3.0truecm
\centerline{Figure~9}\clearpage
\else\figcaption{\figcapnine}\fi
\end{figure}


\begin{figure}
\centerline{\epsfbox{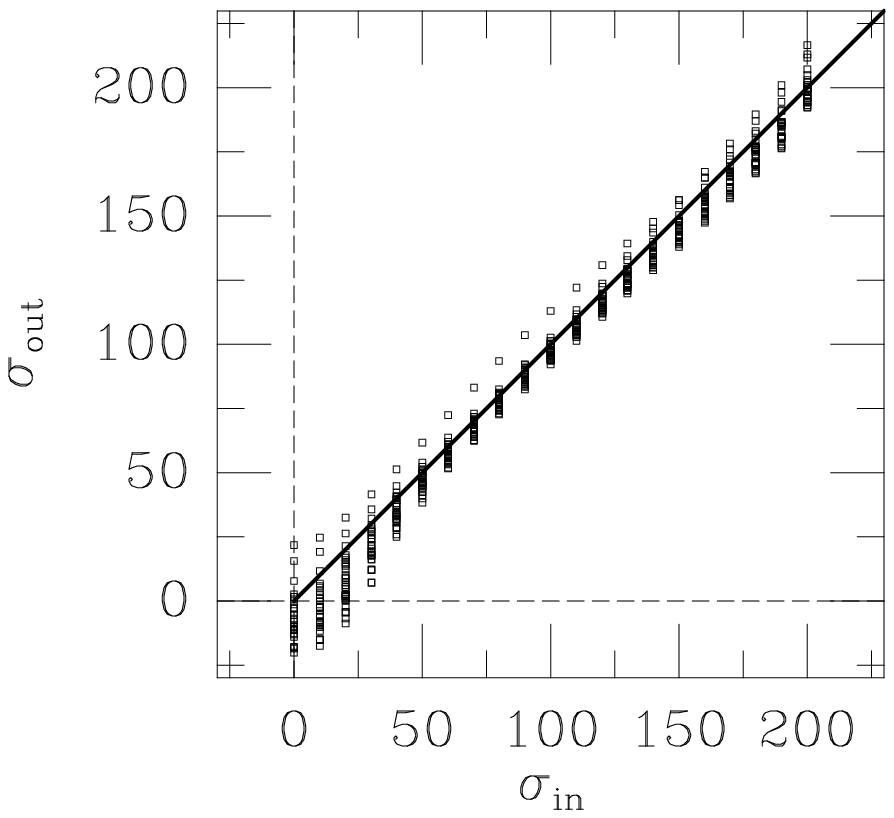}}
\ifsubmode
\vskip3.0truecm
\centerline{Figure~10}\clearpage
\else\figcaption{\figcapten}\fi
\end{figure}

\fi


\clearpage


\begin{deluxetable}{ccccccccr}
\tablecaption{FOS peak-up acquisition strategy\label{t:peakup}}
\tablehead{
\colhead{stage} & \colhead{ap.~name} & 
\multicolumn{2}{c}{nominal size} & \colhead{$N_x$} & \colhead{$N_y$} &
\colhead{$s_x$} & \colhead{$s_y$} & \colhead{$T_{\rm exp}$} \\
 & & \colhead{$x$ (arcsec)} & \colhead{$y$ (arcsec)} & & & 
     \colhead{(arcsec)} & \colhead{(arcsec)} & \colhead{$s$} \\
\colhead{(1)} & \colhead{(2)} & \colhead{(3)} & 
\colhead{(4)} & \colhead{(5)} & \colhead{(6)} &
\colhead{(7)} & \colhead{(8)} & \colhead{(9)} \\
}
\startdata
1 & 4.3        & 3.66  & 1.29  & 2 & 7 & 3.36  & 0.80  &  4.7 \\
2 & 1.0--PAIR  & 0.86  & 0.86  & 6 & 2 & 0.61  & 0.65  &  7.4 \\
3 & 0.5--PAIR  & 0.43  & 0.43  & 3 & 3 & 0.29  & 0.29  &  9.1 \\
4 & 0.25--PAIR & 0.215 & 0.215 & 3 & 3 & 0.143 & 0.143 &  8.5 \\
5 & 0.1--PAIR  & 0.086 & 0.086 & 5 & 5 & 0.043 & 0.043 & 38.7 \\
\enddata
\tablecomments{Column~(1) lists the sequential number of each peak-up stage. 
The name of the aperture is listed in column~(2). Its nominal size is
given in columns~(3) and~(4). The values of $N_x$ and $N_y$ in
columns~(5) and~(6) define the number of grid points for each stage,
while $s_x$ and $s_y$ in columns~(7) and~(8) define the inter-point
spacings. The exposure time per grid point is listed in column~(9).
The first three acquisition stages were performed in the first orbit
of the observations, while the fourth and fifth stage were performed
in the second orbit.}
\end{deluxetable}


\begin{deluxetable}{ccclcccc}
\tablecaption{Spectra: observational setup and intended aperture positions
              \label{t:obsetup}}
\tablehead{
\colhead{ID} & \colhead{HST-ID}  & \colhead{orbit} & 
\colhead{aperture} & \colhead{nominal size} & 
\multicolumn{2}{c}{intended position} & \colhead{$T_{\rm exp}$} \\
 & & & & \colhead{(arcsec)} & \colhead{$x$ (arcsec)} & 
\colhead{$y$ (arcsec)} & \colhead{($s$)} \\
\colhead{(1)} & \colhead{(2)} & \colhead{(3)} & 
\colhead{(4)} & \colhead{(5)} & \colhead{(6)} &
\colhead{(7)} & \colhead{(8)} \\
}
\startdata
S1 & y2uf0107t & 3 & 0.1-PAIR  & $0.086$ & $0.0$ & $ 0.0  $ & 2400 \\
S2 & y2uf0109t & 4 & 0.1-PAIR  & $0.086$ & $0.0$ & $ 0.0  $ & 2400 \\
S3 & y2uf010bt & 5 & 0.1-PAIR  & $0.086$ & $0.0$ & $ 0.086$ & 2400 \\
S4 & y2uf010dt & 6 & 0.1-PAIR  & $0.086$ & $0.0$ & $-0.086$ & 2400 \\
L1 & y2uf010gt & 7 & 0.25-PAIR & $0.215$ & $0.0$ & $ 0.215$ & 1105 \\
L2 & y2uf010ht & 7 & 0.25-PAIR & $0.215$ & $0.0$ & $-0.215$ & 1105 \\
L3 & y2uf010jt & 8 & 0.25-PAIR & $0.215$ & $0.0$ & $ 0.0  $ & 1395 \\
L4 & y2uf010kt & 8 & 0.25-PAIR & $0.215$ & $0.0$ & $ 0.430$ &  800 \\
L5 & y2uf010mt & 9 & 0.25-PAIR & $0.215$ & $0.0$ & $ 0.430$ & 1790 \\
\enddata
\tablecomments{Column~(1) is the label for the spectrum used in 
the remainder of the paper, while column~(2) is the name of the
observation in the HST Data Archive. Column~(3) lists the number of
the orbit in which the spectrum was taken. Column~(4) contains the
name of the square aperture that was used. Column~(5) lists its
nominal size, based on a COSTAR reduction factor of 0.86. Columns~(7)
and~(8) list for each observation the {\it intended\/} aperture
position (Figure~\ref{f:aper_plot}, left panel). The $(x,y)$ coordinate
system is centered on the galaxy, the major axis of which lies along
the $y$-axis. Column~(11) lists the exposure time.}
\end{deluxetable}


\begin{deluxetable}{llcccccc}
\tablecaption{Spectra: actual aperture positions and LSF properties
              \label{t:applaces}}
\tablehead{
\colhead{ID} & \colhead{aperture} & \colhead{calibrated size} & 
\colhead{Intensity} & \multicolumn{2}{c}{actual position} &
\colhead{$\Delta \lambda_{\rm LSF}$} &
\colhead{$\Delta v_{\rm LSF}$} \\
 & & \colhead{(arcsec)} & \colhead{(counts$\>{\rm s}^{-1}$)} & 
     \colhead{$x$ (arcsec)} & \colhead{$y$ (arcsec)}  & 
     \colhead{(\AA)} & \colhead{($\kms$)} \\
\colhead{(1)} & \colhead{(2)} & \colhead{(3)} &
\colhead{(4)} & \colhead{(5)} & \colhead{(6)} &
\colhead{(7)} & \colhead{(8)} \\
}
\startdata
S1 & 0.1-PAIR  & 0.068 &  218.1 & $-0.010$ & $ 0.015$ & $ 0.009$ & $ 0.54$ \\
S2 & 0.1-PAIR  & 0.068 &  203.6 & $ 0.000$ & $ 0.040$ & $ 0.000$ & $ 0.00$ \\
S3 & 0.1-PAIR  & 0.068 &  139.7 & $ 0.000$ & $ 0.126$ & $ 0.000$ & $ 0.00$ \\
S4 & 0.1-PAIR  & 0.068 &  199.2 & $ 0.000$ & $-0.046$ & $ 0.000$ & $ 0.00$ \\
L1 & 0.25-PAIR & 0.191 &  686.0 & $-0.047$ & $ 0.304$ & $ 0.030$ & $ 1.72$ \\
L2 & 0.25-PAIR & 0.191 & 1133.7 & $-0.047$ & $-0.126$ & $ 0.087$ & $ 5.03$ \\
L3 & 0.25-PAIR & 0.191 & 1269.3 & $-0.047$ & $ 0.089$ & $ 0.110$ & $ 6.41$ \\
L4 & 0.25-PAIR & 0.191 &  435.5 & $-0.047$ & $ 0.519$ & $ 0.013$ & $ 0.78$ \\
L5 & 0.25-PAIR & 0.191 &  427.7 & $-0.024$ & $ 0.525$ & $ 0.007$ & $ 0.39$ \\
\enddata
\tablecomments{Column~(1) is the label of the spectrum, while column~(2)
is the square aperture with which it was obtained. Column~(3) lists
the size of the aperture as derived in Appendix~\ref{s:AppA} by
fitting to calibration observations of a star. Column~(4) lists the
observed intensity in counts per second integrated over the wavelength
range covered by the grating. The formal errors in these intensities
due to Poisson statistics are $\leq 1.0$ for all observations.
Columns~(5) and~(6) give the position of the aperture for each
observation (Figure~\ref{f:aper_plot}, right panel), inferred from the
observed intensity as described in the text. The $(x,y)$ coordinate
system is centered on the galaxy, the major axis of which lies along
the $y$-axis. Column~(7) lists the mean of the LSF in {\AA},
calculated as described in Section~\ref{s:lsf}. Column~(8) lists the
corresponding velocity shift at 5170{\AA}.}
\end{deluxetable}


\begin{deluxetable}{lrrrrrr}
\tablecaption{Line strengths and kinematics
              \label{t:kinresults}}
\tablehead{
\colhead{ID} & \colhead{$\gamma$} & \colhead{$\Delta \gamma$} & 
\colhead{$V$} & \colhead{$\Delta V$} &
\colhead{$\sigma$} & \colhead{$\Delta \sigma$} \\
 & & & \colhead{($\kms$)} & \colhead{($\kms$)} & \colhead{($\kms$)} &
       \colhead{($\kms$)} \\ }
\startdata
S1  & 0.941 & 0.038 & $  9.1$ & 10.0 & 156.4     & 10.2 \\
S2  & 0.957 & 0.033 & $ 24.5$ &  6.4 &  95.6     &  9.8 \\
S3  & 1.014 & 0.046 & $ 33.5$ &  9.7 & 111.9     & 11.7 \\
S4  & 1.049 & 0.039 & $-28.1$ &  8.6 & 127.0     &  9.6 \\
L1  & 0.927 & 0.027 & $ 39.9$ &  5.3 &  78.3     & 11.0 \\
L2  & 0.989 & 0.022 & $-35.9$ &  4.3 &  83.4     &  9.3 \\
L3  & 1.104 & 0.022 & $ 19.3$ &  4.8 & 124.2     &  7.0 \\
L45 & 0.973 & 0.022 & $ 45.1$ &  4.3 &  70.0$^*$ &  5.0 \\
\enddata
\tablecomments{Results of the kinematical analysis of the galaxy 
spectra, as discussed in Section~\ref{s:kinematics}. The aperture
positions for the observations are listed in
Table~\ref{t:applaces}.\hfill\break
\null\qquad$^*\>$The dispersion listed for L45 is the one measured 
from ground-based data at the same position, as discussed in the text.}
\end{deluxetable}


\begin{deluxetable}{ccccllc}
\tablecaption{PSF parameters and aperture sizes\label{t:psfapfit}}
\tablehead{
\multicolumn{3}{c}{Gaussian PSF parameters} & \hspace{2truecm} &
\multicolumn{3}{c}{Aperture sizes} \\
\colhead{$i$}  & \colhead{$\gamma_i$} & \colhead{$\sigma_i$} & \hspace{2truecm} &
\colhead{Name} & \colhead{shape}      & \colhead{size/diameter} \\
               &                      & \colhead{(arcsec)}   & \hspace{2truecm} &
               &                      & \colhead{(arcsec)}      \\
}
\startdata
1 & 0.6198 & 0.0327 & \hspace{2truecm} & 0.1-PAIR  (upper) & square   & 0.068 \\
2 & 0.3012 & 0.1043 & \hspace{2truecm} & 0.25-PAIR (upper) & square   & 0.191 \\
3 & 0.0790 & 0.6387 & \hspace{2truecm} & 1.0               & circular & 0.825 \\
\enddata
\tablecomments{The listed parameters yield the best fit (solid curves 
in Figure~\ref{f:transmis}) to the calibration data of Evans (1995a),
using the model described in the text.}
\end{deluxetable}


\end{document}